\documentclass[aps,preprint]{revtex4}
\usepackage{graphicx}
\usepackage{bm}
\begin{document}
 \date{\today}
\title{Correlated behavior of conductance and phase rigidity in the transition
 from the weak-coupling to the strong-coupling regime 
}
\author{E.N. Bulgakov$^{1,2}$, I. Rotter$^1$ and A. F. Sadreev$^{1,2}$}
\affiliation {$^1$ Max Planck Institute for the Physics of Complex
Systems, D-01187 Dresden, Germany }
\affiliation{$^2$  Kirensky
Institute of Physics, 660036, Krasnoyarsk, Russia}

\begin{abstract}
We study the transmission through different small systems as a
function of the coupling strength $v$ to the two attached leads.
The leads are identical with only one propagating mode $\xi^E_C$ in each
of them. Besides the conductance $G$, we calculate the
phase rigidity $\rho$ of the scattering wave function 
$\Psi^E_C$ in the interior of the system.
Most interesting results are obtained in the regime of strongly overlapping 
resonance states where  the crossover from staying to traveling modes takes
place. The crossover is characterized by collective effects.
Here, the conductance is plateau-like enhanced in some  energy regions of 
finite length while corridors with zero transmission
(total reflection) appear in other   
energy regions. This transmission picture depends only weakly on the spectrum 
of the closed system. It is caused by the alignment of some resonance states
of the system  with the propagating modes $\xi^E_C$ in the leads.
The alignment of  resonance states takes place stepwise
by resonance trapping, i.e. it is accompanied by the decoupling
of other resonance states from the continuum of propagating modes.
This process is quantitatively described by the 
phase rigidity $\rho$ of the scattering wave function.
Averaged over energy in the considered energy window, 
$\langle G\rangle$ is correlated with $1-\langle\rho\rangle$. 
In the regime of strong coupling, only
two  short-lived resonance states survive each aligned with one of the 
channel wave functions $\xi^E_C$. They
may be identified with traveling modes through the system.
The remaining $M-2$ trapped narrow resonance states are well separated
from one another.  
\end{abstract}

\maketitle

\section{Introduction}

Resonance phenomena in quantum systems at very low   and at 
very high level density are well understood. In the first case, isolated
resonances can be seen in the cross section whose decay widths  
are smaller than the distance between them. They are described well by
means of the  discrete states of the closed system. In the second case 
narrow resonances appear, superposed on a smooth background. Here,  
the resonance states are usually described on the basis of statistical
assumptions while the background is represented by the so-called optical $S$
matrix \cite{feshbach}. Much less understood   is   the transition between
these two borderline cases. In this crossover regime, the resonances overlap, 
and a full (standard) statistical description is, generally, not  justified. 
For the results of numerical studies  for a realistic system
see, e.g., Ref. \cite{drozdz}.
Characteristic features of the scenario with overlapping resonances are the 
avoided crossings of the resonance states 
in the complex plane which may change considerably 
the spectroscopic properties (such as positions and widths)
of the states \cite{ro91rep}. In order to describe these changes under
realistic conditions (especially in the few-channel case), 
both the real {\it and} imaginary
parts of the coupling via the continuum have to be taken into account.
In the standard statistical approach, only the real part is considered. 
In the recent paper \cite{izzel}, only the imaginary part of the coupling 
via the continuum is taken into account. Here, serious
deviations from the standard statistical approach are found. 
Strong deviations are found also when both the real and imaginary parts are
considered \cite{drozdz}. Generally, the phases of the wave functions are 
not rigid in the regime of overlapping resonances
as shown analytically in Ref. \cite{braro}. 

Mathematically, an open quantum system 
can be described best by a non-hermitian Hamilton operator that accounts not
only for the energies but also for the widths (lifetimes) of the resonance
states. Spectroscopic studies on the basis of a realistic
non-hermitian Hamilton operator (with complex coupling term) are
performed up to now either by using the method of complex scaling  
\cite{moisrep} or the method of the Feshbach projection operator 
technique \cite{ro91rep}. The last method is applied also to the 
transmission through small devices. The theory for this case, 
presented in Ref. \cite{saro},  is equivalent to the tight-binding 
lattice model \cite{datta}.
In the numerical results, long-range correlations between the 
individual resonance states have been observed. 

Most theoretical studies with non-hermitian Hamilton operators are based on 
random matrices. Usually, the Pandey-Mehta Hamilton operator 
$H(\alpha)=H_0+\alpha H_1$ is used where $H_0$ and
$H_1$ are real and complex random hermitian matrices, respectively, and
$\alpha$ is a crossover parameter \cite{pandey}. Recent theoretical studies 
have shown that long-range correlations in the eigenfunctions of $H(\alpha)$ 
appear also in these studies \cite{brouwer2}. 
The  theoretical results  are in  good agreement with experimental ones 
obtained in an open microwave billiard   \cite{brouwerkuhl}.

In order to describe the intensity fluctuations
generated by a monochromatic source in a disordered cavity coupled to the
environment, the idea of {\it standing} waves in  almost closed systems and 
{\it traveling} waves in open systems is considered in Ref. \cite{shapiro}. 
In this paper, the two borderline cases are related to one another by 
an interpolation procedure.

Another characteristic feature of overlapping resonances is
the fluctuation picture of the cross section. The 
autocorrelation function shows   Ericson
fluctuations  \cite{ericson} studied experimentally first in 
nuclear reaction cross sections \cite{brentano}. 
Recently they are studied experimentally with high resolution in the 
photoionization of rubidium atoms in crossed magnetic and electric fields
\cite{stania}. The experimental results 
obtained in this study proved not only the universality of the Ericson
fluctuations. They showed
an unexpected  high reproducibility of the results.
This experimental result does not coincide with the usual understanding of
the nature of Ericson fluctuations.

In recent theoretical studies on Ericson fluctuations, 
the importance of interferences between
different resonance states is discussed. In Ref. \cite{mois} 
it is presumed that Ericson fluctuations do not result from interferences, but 
may be considered as a collective coherent resonance phenomenon. 
In the literature, this collective phenomenon is called 
resonance trapping.  It is studied  
theoretically as well as  experimentally \cite{perssonexp} 
in many papers, see the reviews \cite{ro91rep}. It is a collective effect
caused by the decoupling of most of the resonance 
states from the continuum   at large coupling strength 
between system and environment while only a few states become short-lived.  
This interpretation of Ericson fluctuations is called 
misinterpretation in Ref. \cite{kun} due to the  result 
obtained by the author, that the fluctuations originate from interferences 
between random-partial-width amplitudes in the regime of strongly overlapping 
resonances. 

It is the aim of the present paper to study
the transmission through microwave cavities as well as through toy-model 
systems with a small number of sites (corresponding to a small number 
of levels) in the two-channel case (one in each of the attached
leads) without using any statistical assumptions. The mathematical basis 
of our study is the Feshbach projection operator \cite{feshbach} formalism, 
i.e. the non-hermitian effective Hamilton operator
$H_{\rm eff}$ (see Sect. II) and its relation to the Green function  
$G_{\rm eff} = (E-H_{\rm eff})^{-1}$. 
The numerical results for microwave cavities (Sect. III)
as well as for one-dimensional and two-dimensional toy-model systems 
with different positions of the attached wires
(Sect IV) are obtained by using the tight-binding lattice Green
function method \cite{datta}  providing accurate results
for sufficiently large numerical grids.
We turn our attention especially to the appearance of coherent collective
effects and long-range correlations in the 
crossover regime from standing to traveling waves, i.e. in the
regime of overlapping resonances. Furthermore (Sect. V), we show the
appearance of Ericson-like fluctuations. We are interested in the 
structure of the autocorrelation function for different values of the 
displacement and for different degrees of resonance overlapping.
We comment also on the reproducibility of the results by 
changing the geometry of the system.
The results are discussed and summarized in the last section.

\section{Spectroscopic values of an open quantum billiard}

\subsection{Non-hermitian Hamilton operator of the open system}

In an exact description of resonance phenomena by using the Feshbach
projection operator technique \cite{feshbach},
the non-hermitian effective Hamilton operator
%------------------------------------------------------------------------(1)
\begin{equation}
\label{Heffgen}
H_{\rm eff}=H_B+\sum_{C=L,R}
V_{BC}\frac{1}{E^{+}-H_C}V_{CB}
\end{equation}
appears
where $H_B$ is the hermitian Hamilton operator of the corresponding closed
system with $M$ discrete states, $V_{BC}$ is the symmetrical
coupling operator between system  and environment, 
$H_C$ is the hermitian Hamilton operator describing the
propagating modes in the environment and $C$ stands 
for the continuum of propagating modes. In the case of a quantum
cavity, $C$ consists of the waves in the right ($R$) and left ($L$) 
leads attached to the cavity \cite{saro}.  In this representation,
the resonance picture is determined by the eigenvalues
$z_\lambda$ and eigenfunctions $\phi_\lambda$ of $H_{\rm eff}$,
Eq. (\ref{Heffgen}), which differ, generally, from those of $H_B$
not only by Im$(z_\lambda) \ne 0$. Also Re$(z_\lambda)$ differs from the
(real) eigenvalue of $H_B$ by the contribution arising from the principal value
integral of the second term of $H_{\rm eff}$.  

The $M$ eigenvalues $z_\lambda$ of $H_{\rm eff}$ 
are complex and energy dependent. They
provide the positions $E_\lambda$ and widths $\Gamma_\lambda$ of the 
resonance states by solving the fixed-point equations
$E_\lambda  = 
{\rm Re}\big(z_\lambda\big)\big|_{E=E_\lambda}$ and defining
$\Gamma_\lambda  =  
- 2 \; {\rm Im}\big(z_\lambda\big)\big|_{E=E_\lambda}$.
These values correspond to the poles of the $S$ matrix. 
However,
for the $S$ matrix describing a physical scattering process at the real energy
$E$ of the system 
(without any poles), the solutions of the fixed-point equations are 
relevant only for narrow resonances. Generally, 
the  eigenvalues $z_\lambda$ with 
their full energy dependence are involved in the $S$ matrix \cite{ro91rep}.  

Also the eigenfunctions $\phi_\lambda$ of $H_{\rm eff}$ are  complex and 
energy dependent. The $\phi_\lambda$ are biorthogonal.
Due to the symmetry of $H_{\rm eff}$, it holds \cite{ro91rep}, 
%------------------------------------------------------------------------(2)
\begin{equation}
\langle \phi_\lambda^*|\phi_{\lambda '} 
\rangle = \delta_{\lambda,\lambda '} \; .
\label{biorth}
\end{equation}
As a consequence, 
$\langle \phi_\lambda | \phi_\lambda \rangle = A_\lambda\ge 1 $
and $A_\lambda \to \infty$ with approaching the branch point 
at which two eigenvalues coalesce, $z_\lambda
= z_{\lambda '}$ ~\cite{rstrans,rstopol,brsphas}. 

Considering the coupling strength $v$ as control
parameter, the phenomenon of resonance trapping
\cite{klro,ro91rep} appears, i.e. the wave functions of a few resonance states 
align each with one of the channel 
wave functions while the other resonance states decouple from the continuum
(become trapped). The resonance trapping occurs hierarchically 
\cite{isrodi,mudiisro}.
Some years ago, the phenomenon of resonance trapping is studied
theoretically  \cite{seba}  and even proven 
experimentally \cite{perssonexp} on microwave cavities. 

The eigenfunctions $\phi_\lambda$ of $H_{\rm eff}$ can be represented 
in the set $\{\phi_\lambda^{B}\}$ of eigenfunctions 
of the Hamilton operator $H_B$ of the corresponding closed
system, 
%------------------------------------------------------------------------(3)
\begin{eqnarray}
\phi_\lambda = \sum_{\lambda '}^M d_{\lambda \lambda '} ~\phi_{\lambda '}^B
\label{wfrepr}
\end{eqnarray}
with complex coefficients $d_{\lambda \lambda '}$ normalized according to 
$|d_{\lambda \lambda '}|^2 /
\sum_{\lambda " =1}^M |d_{\lambda '\lambda "}|^2$.

\subsection{Scattering wave function and transmission}
 
The eigenfunctions $\phi_\lambda$ of $H_{\rm eff}$ are part of
the total scattering wave function $\Psi_C^E$ that
is solution of the Schr\"odinger equation $(H-E)\Psi^E_C=0 $ 
in the total function space consisting of the discrete states 
at the energies $E_\lambda^B$ of the closed quantum billiard 
and of the propagating modes $\xi^E_C$ in the attached leads.
Due to the coupling between the two subspaces, the
discrete states turn over into resonance states at the energies $E_\lambda$
of the open billiard. $H$ is hermitian. The $\Psi_C^E$ read \cite{ro91rep}
%------------------------------------------------------------------------(4)
\begin{equation}
\Psi_C^E = \xi^E_C + \sum_{\lambda} 
\Omega^C_\lambda  \,
\frac{\langle\phi_\lambda^* |V| \xi^E_C\rangle}{E-z_\lambda} 
\label{total}
\end{equation}
where $\Omega^C_\lambda  =
\big[ 1+(E^{+}-H_C)^{-1} V\big]\,\phi_\lambda  $
is the wave function of the resonance state $\lambda$. 
According to (\ref{total}), the eigenfunctions $\phi_\lambda$ of the 
effective Hamilton operator $H_{\rm eff}$ give an essential
contribution to the scattering wave function $\Psi^E_C$ in the interior 
of the cavity, $\Psi^E_C \to \hat \Psi^E_C$, with
%------------------------------------------------------------------------(5)
\begin{eqnarray}
\hat\Psi_C^E = \sum_\lambda c_{\lambda E}\, \phi_\lambda  \, ; 
\quad \; c_{\lambda E} =
\frac{\langle \phi_\lambda^* |V| \xi^E_C\rangle}{E-z_\lambda } \; .
\label{total1}
\end{eqnarray}
The coupling coefficients $\langle\phi_\lambda^* |V| \xi^E_C\rangle $
are, generally,  complex and energy dependent due to the unitarity of the
$S$ matrix \cite{ro03}. This energy dependence becomes important only for
overlapping resonances. Here one of the resonances may appear as some
"background" onto which a neighboring resonance is superposed \cite{marost4}.

In Refs. \cite{brouwer1,brouwer2}, the phase rigidity $|\rho|^2$ with
%------------------------------------------------------------------------(6)
\begin{eqnarray}
%\rho_{\rm br} =\frac{\int dr ~\hat\Psi(r)^2}{\int dr ~|\hat\Psi(r)|^2}
\rho =\frac{\int dr ~\hat\Psi(r)^2}{\int dr ~|\hat\Psi(r)|^2}
= e^{2i\theta} 
\frac{\int dr ~([{\rm Re}\hat\Psi(r)]^2 - [{\rm Im}\hat\Psi(r)]^2)}
{\int dr ~([{\rm Re}\hat\Psi(r)]^2 + [{\rm Im}\hat\Psi(r)]^2)}
\label{rig2}
\end{eqnarray}
defined at the energy $E$, is introduced 
in order to describe phenomenologically the influence of some impurity  
onto the phase
of the scattering wave function $\hat\Psi^E_C$. We use this expression 
in order to get information on the phases of the $\hat\Psi^E_C$
in the regime of resonance overlapping.

The amplitude of the transmission through a quantum dot is \cite{saro}
%--------------------------------------------------------------------(7)
\begin{equation}
t=-2\pi
i\sum_{\lambda}\frac{\langle \xi^E_L|V|\phi_\lambda\rangle
\langle\phi_\lambda^*|V|\xi^E_R\rangle }{E-z_{\lambda}} \; . 
\label{trHeff}
\end{equation}
The eigenvalues $z_\lambda$ and eigenfunctions $\phi_\lambda$ of $H_{\rm eff}$
are involved in  (\ref{trHeff}) with their full energy dependence 
\cite{ro91rep}.
According to (\ref{trHeff}), the transmission is  resonant in relation 
to the real part of the eigenvalues of
$H_{\rm eff}$. 
The transmission amplitude (\ref{trHeff}) can be rewritten
by means of the scattering wave function (\ref{total1}),
%-------------------------------------------------------------------(8)
\begin{eqnarray}
t = - 2\pi i
~\langle \xi^E_C|V|\hat\Psi^E_C\rangle 
\label{tr}
\end{eqnarray}
with $\hat\Psi^E_R$ being complex, in general.
The advantage of this representation consists in the fact that it does not
suggest the existence of Breit-Wigner peaks in the transmission probability. 
Quite the contrary, the transmission is determined by the degree of alignment
of the wave function $\hat \Psi_C^E$ with the propagating modes 
$\xi^E_C$ in the leads, i.e. by the value 
$\langle \xi^E_C|V|\hat\Psi^E_C\rangle $. This alignment is important in the
regime of overlapping resonances where the identification of single resonances
and resonance states ceases to be meaningful. The expressions(\ref{trHeff})
and (\ref{tr})  are fully equivalent.

\subsection{Correlation measures}

The redistribution processes taking place inside the system under the 
influence of the coupling to the continuum can be characterized by
the number $N_\lambda^p$ of principal components of the
eigenfunctions $\phi_\lambda$. Using  (\ref{wfrepr}) 
this number reads \cite{jung}
%---------------------------------------------------------------------(9)
\begin{eqnarray}
N_\lambda^p=\frac{1}{M\sum_{\lambda ' =1}^M |b_{\lambda \lambda '}|^4}
\quad ; \qquad |b_{\lambda \lambda '}|^2=\frac{|d_{\lambda \lambda '}|^2}{
\sum_{\lambda " =1}^M |d_{\lambda ' \lambda "}|^2} \; .  
\label{pricom1}
\end{eqnarray}
In the limit of equal mixing of the state $\lambda$  with all states 
$\lambda '$, one gets $b_{\lambda \lambda '} =1/\sqrt{M} $ for all $\lambda '$
and $N_\lambda^p=1$. In the opposite case, we have $b_{\lambda \lambda '}
= \delta_{\lambda, \lambda '}$ and $N_\lambda^p=1/M$. In the first case 
the collectivity of the state $\lambda$ is large \cite{jung,izzel}.
Almost all components of the wave function have the same sign 
because  the wave function of the  state $\lambda$ is aligned
to the scattering wave function $\xi^E_C$ of the channel $C$
at the cost of the other resonance states trapped by it. 
In the second case the wave functions
$\lambda$ are not at all aligned  to the $\xi^E_C$, and the signs of their
different components are not the same. For details of the phenomenon of
alignment see Ref. \cite{braro}. Generally, $1/M \le N_\lambda^p \le 1$.

Due to (\ref{total1}), the  properties of the scattering wave 
function $\hat\Psi_C^E$ are determined by the  properties
of all the wave functions $\phi_\lambda$. 
Using the representation (\ref{wfrepr}), we can write $\hat\Psi^E_C = 
\sum_\lambda \langle \phi_\lambda^B|\hat\Psi^E_C\rangle ~\phi_\lambda^B$
and define 
%-------------------------------------------------------------------(10)
\begin{eqnarray}
N_E= 
\frac{1}{\sum_{\lambda =1}^M |\langle \phi_\lambda^B|\hat\Psi^E_C\rangle|^4} 
\; .
\label{pricom2}
\end{eqnarray}
The normalization has to be done separately
at every energy $E$ and coupling strength $v$.
The value $N_E$ is a certain analogon to (\ref{pricom1}),
but for the representation of the scattering wave function $\hat\Psi^E_C$
in the set of eigenfunctions $\phi_\lambda^B$ of the Hamilton operator $H_B$.
Usually, $N_E$ is called   inverse participation ratio.
Further, we define the analogous expression
%-------------------------------------------------------------------(11)
\begin{eqnarray}
N_x= \frac{1}{|\langle \overline{x}|\hat\Psi^E_C\rangle|^4} 
\label{pricom3}
\end{eqnarray}
with $\hat\Psi^E_C = 
\langle \overline{x}|\hat\Psi^E_C\rangle ~\phi_\lambda^B$.

Another measure for coherence 
is the autocorrelation function of the fluctuation
$\delta \sigma = \sigma - \overline{\sigma}$ of the cross section $\sigma$,
%-------------------------------------------------------------------(12)
\begin{eqnarray}
C(\varepsilon)
= \frac{1}{\overline{\sigma}^2 |I|} \int_I
\delta \sigma(E+\varepsilon)\delta \sigma(E)dE \; . 
\label{eric1}
\end{eqnarray}
Here, $I$ is the same energy interval used to
determine $\overline{\sigma}$ and $\delta \sigma = \sigma -
\overline{\sigma}$. The Lorentzian shape of $C(\varepsilon)$ 
for small $\varepsilon$ is characteristic of Ericson fluctuations 
\cite{ericson}. 
In the overlapping regime, the width of the Lorentzian 
is large compared to the average spacing of the resonance states.
Such a result has been found in many experimental data, above all 
in the analysis of nuclear reaction cross sections, e.g. \cite{brentano}.
It has been found recently also in high-resolution 
experimental data on the photoionization of atoms \cite{stania}.
In the present paper, we will analyse the fluctuation picture of the 
transmission 
and denote the corresponding autocorrelation function by $C_G(\varepsilon)$.

\section{Numerical results: transmission
through quantum billiards, energy-averaged phase rigidity 
and collectivity}

The numerical calculations are performed for chaotic  quantum billiards
of different type in the tight-binding lattice model, for details see 
Datta \cite{datta} and \cite{saro}.
The phase rigidity $\rho$ is calculated in the representation (\ref{rig2}).
The coupling strength $v$ between cavity and attached leads is the same
for both leads. It is  varied by means of $v\equiv
\kappa/\kappa_0$ where $\kappa_0=1$ is 
the hopping matrix element inside the cavity as well as inside the lead 
while $\kappa$ is the hopping matrix element between cavity and lead. 
The cavities are small and the number of channels is one in each of the two 
attached identical  leads. We trace  
transmission and phase rigidity (averaged over the whole energy window
between 11 and 34, in units of the widths of the leads)
as a function of the coupling strength $v$ between cavity and attached leads.

\begin{figure}[t]
\includegraphics[width=.57\textwidth]{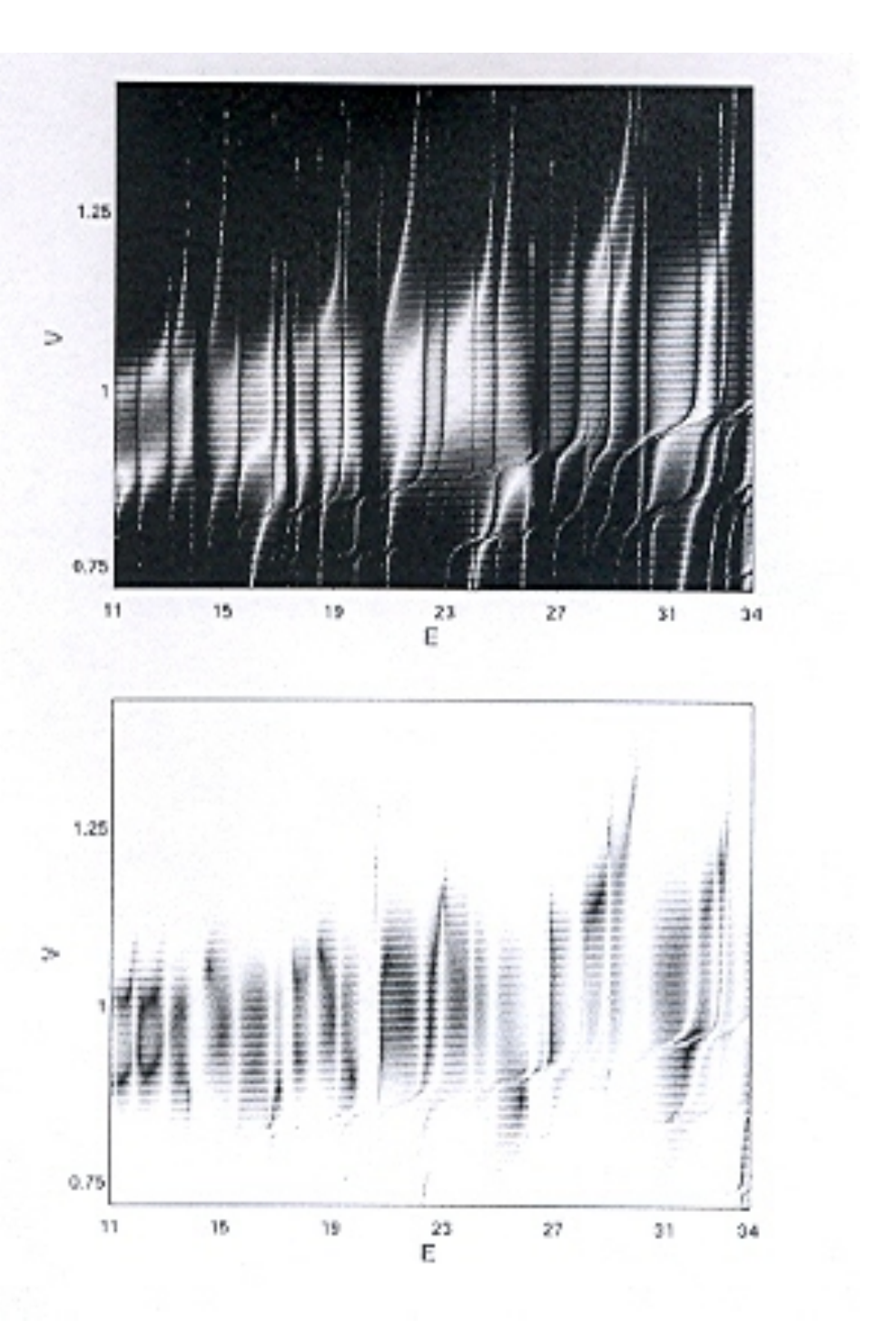}\\
\caption{\small
Conductance $G=|t|^2$ (top) and phase rigidity $|\rho|^2$ (bottom)
for a Sinai billiard as a function of energy $E$ and
coupling strength $v$ between billiard and attached leads \cite{comm}. 
The transmission and the phase rigidity vary between 1 (white) and 0 (black).
The shape of the billiard is shown in Figs. \ref{figwf}(a,b): 
size $x=3, ~y=7$, radius  $R=1.5$, in units of the width
of the leads. 
The calculations are performed in the tight-binding lattice model
\cite{datta}.  
}
\label{figsin1}
\end{figure}

\begin{figure}[t]
\includegraphics[width=.5\textwidth]{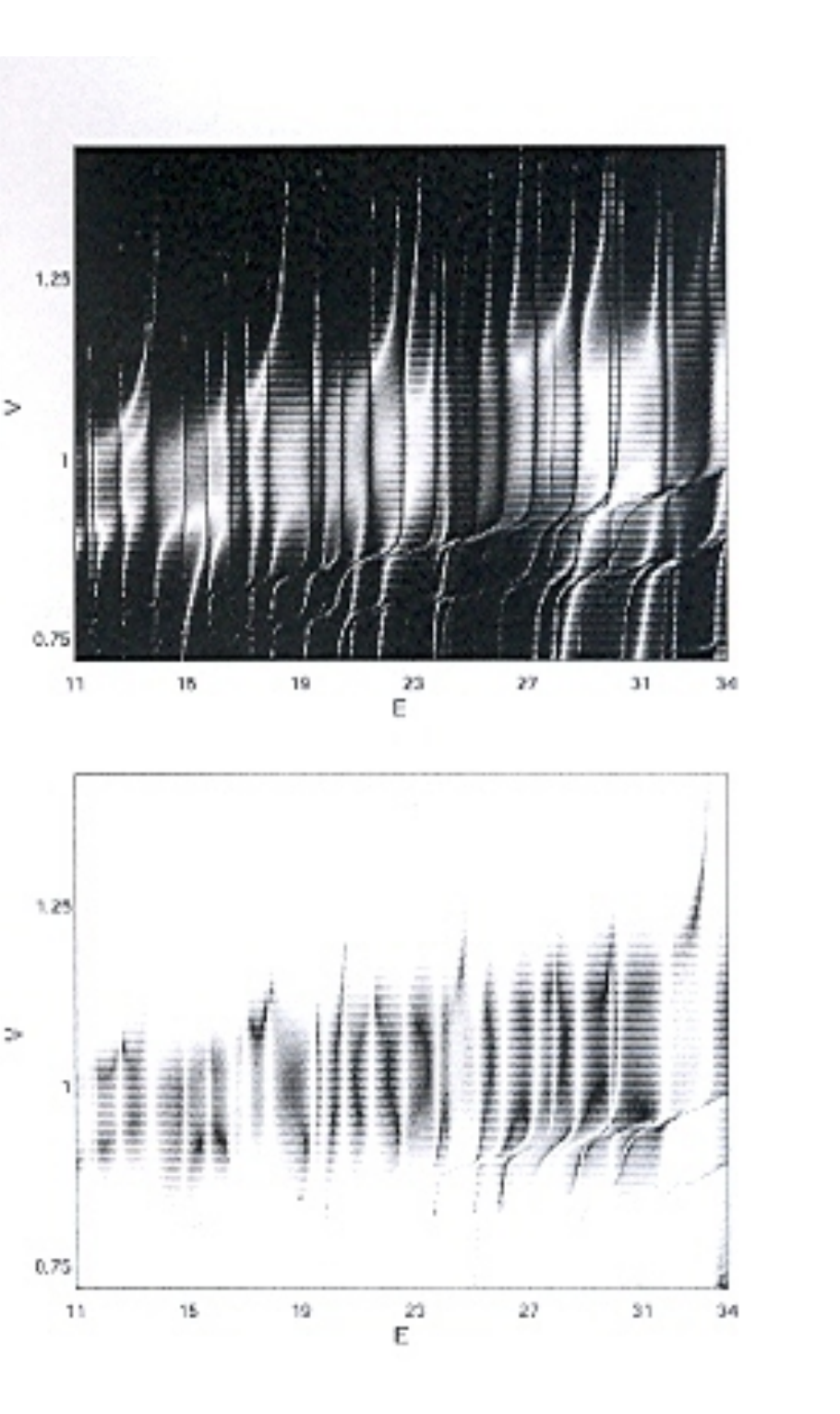}\\
\caption{ \small
The same as Fig. \ref{figsin1} but $x=4, y=5$. The shape of the
billiard is shown in Figs. \ref{figwf}(c,d). 
}
\label{figsin2}
\end{figure}

\begin{figure}[t]
\includegraphics[width=.5\textwidth]{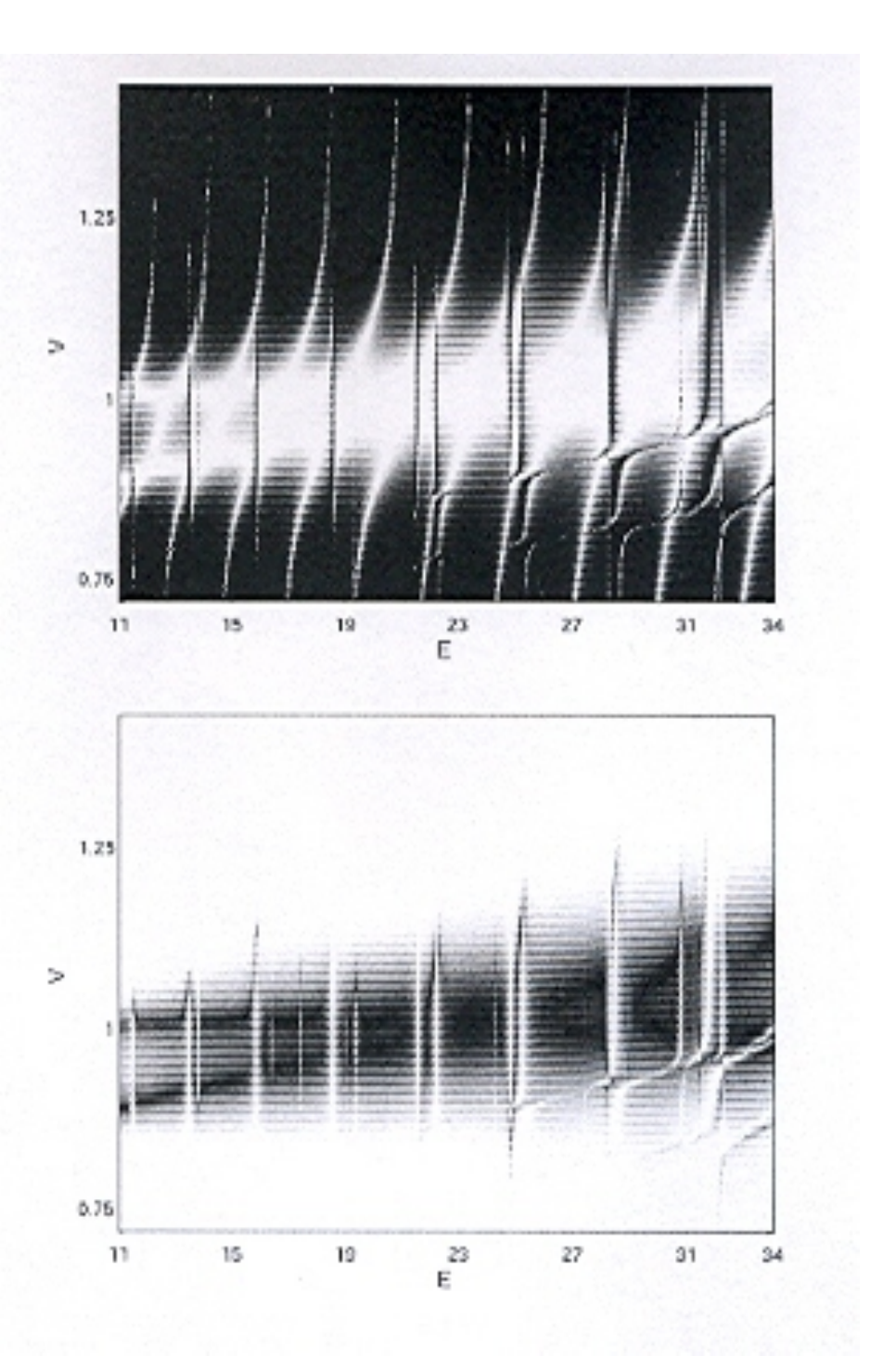} \\
\caption{\small
The same as Fig. \ref{figsin1} but for a billiard 
of Bunimovich type to which the leads are attached in such a manner that 
transmission via whispering gallery modes is supported. The shape of the
billiard is shown in Figs. \ref{figwf}(e,f): Radius $R=3$ and distance
$D=2$ between the centers, in units of the width of the leads. 
}
\label{figbun}
\end{figure}

\begin{figure}[t]
\includegraphics[width=.45\textwidth]{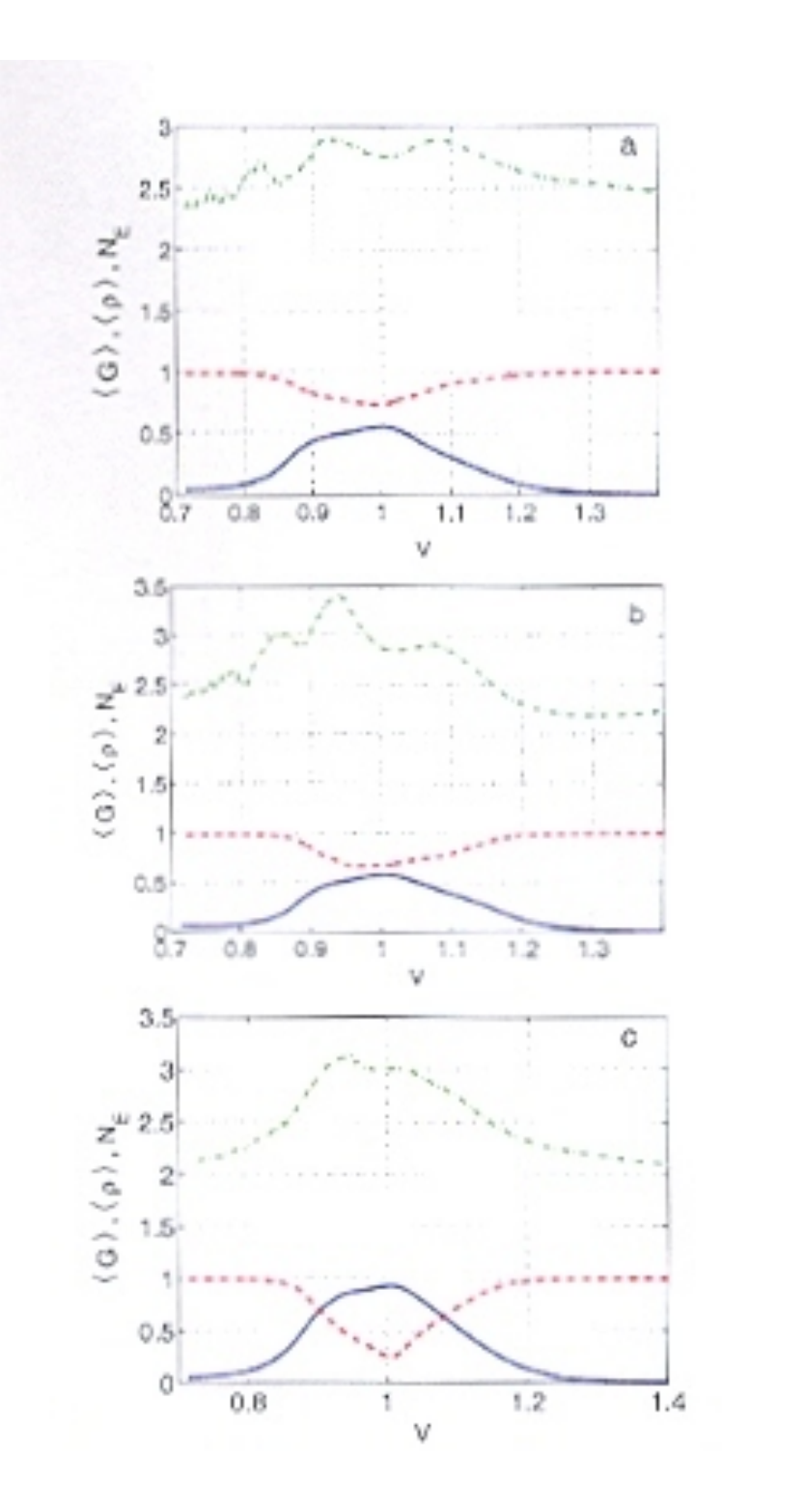}
\caption{\small (color online)
Conductance $\langle G \rangle \, = \, \langle |t|^2 \rangle $ 
(blue full line), energy-averaged  
phase rigidity $\langle |\rho|^2\rangle$ (red dashed line)
and inverse participation ratio $N_E$  of the scattering wave function 
$\Psi^E_C$ (green dash-dotted line),
averaged over the whole energy window ($11 \le E \le 34$), as a function
of the coupling strength $v$, ~\cite{comm}.
(a) Sinai billiard with $x=3, y=7, R=1.5$ in units of the width of
the leads [Figs. \ref{figwf}(a,b)]; (b) the same as (a) but
$x=4, y=5$ [Figs. \ref{figwf}(c,d)]; (c) Bunimovich billiard with  
transmission through whispering gallery modes,
$R=3, D=2$ in units of the width of the leads [Figs. \ref{figwf}(e,f)].
The calculations are performed in the tight-binding lattice model
\cite{datta}.  
}
\label{figcorr}
\end{figure}

\begin{figure}[t]
\includegraphics[width=.58\textwidth]{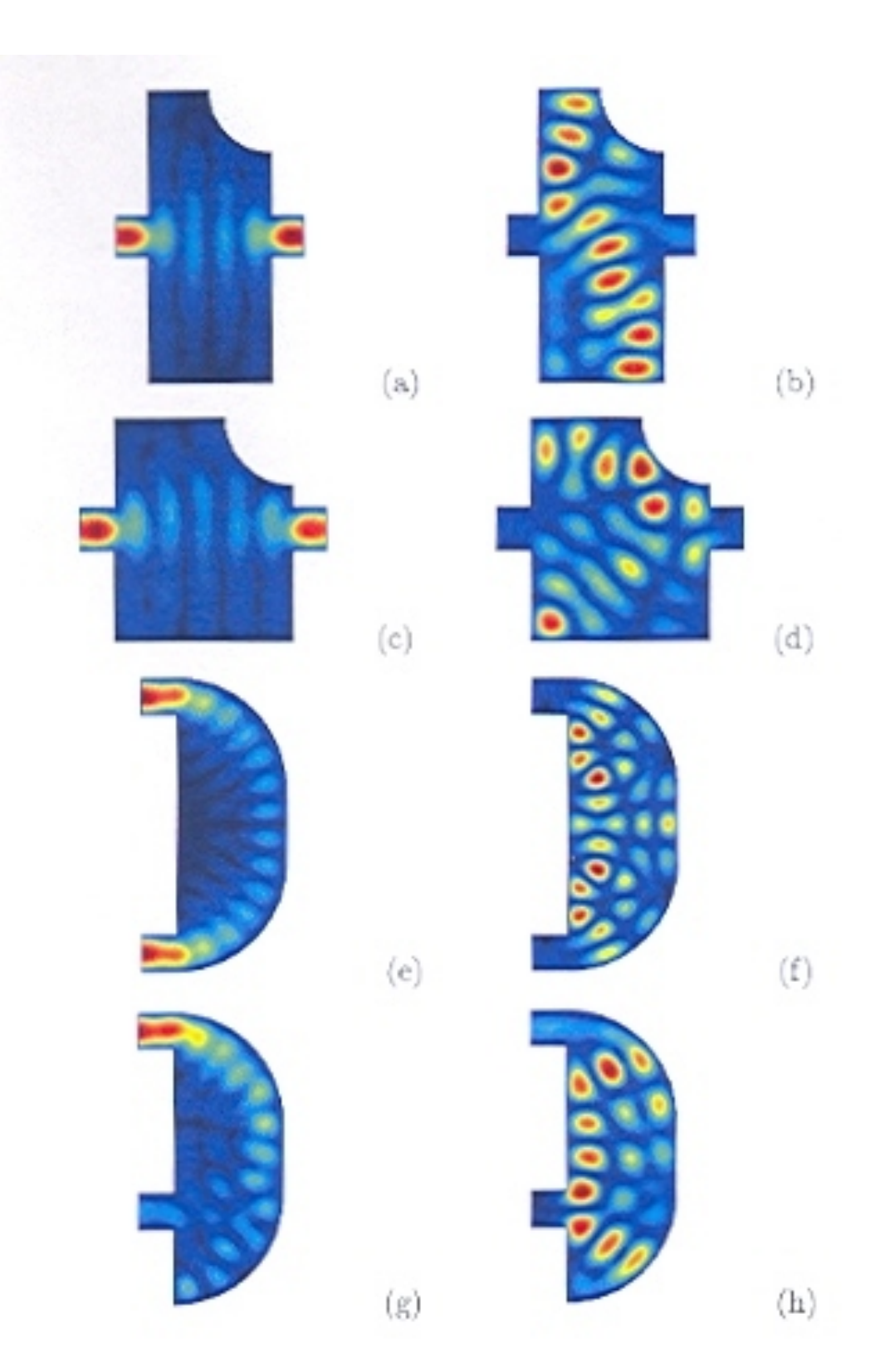}
%\includegraphics[width=.254\textwidth]{fig5b.eps}(b)\\
%\hspace*{.25cm}
%\includegraphics[width=.194\textwidth]{fig5c.eps} \hspace*{.25cm}
%(c)\hspace*{0.8cm}
%\includegraphics[width=.194\textwidth]{fig5d.eps}\hspace*{.25cm}(d)\\
%\includegraphics[width=.254\textwidth]{fig5e.eps}(e)
%\includegraphics[width=.254\textwidth]{fig5f.eps}(f)\\
%\includegraphics[width=.254\textwidth]{fig5g.eps}(g)
%\includegraphics[width=.254\textwidth]{fig5h.eps}(h)
\caption{\small (color online)
Wave functions $\phi_\lambda$ of some short-lived and long-lived resonance 
states in the Sinai billiard (a,b) with $x=3, y=7$; (c,d) with $x=4, y=5$,
and in the Bunimovich billiard (e,f) with leads attached for support of
transmission via whispering gallery modes 
and (g,h) without support of whispering gallery modes.
The corresponding eigenvalues $z_\lambda$ are given in Table \ref{tab}.
The short-lived states (a,c,e,g) are localized in space and de-localized in
energy while the long-lived states (b,d,f,h)
are de-localized in space and localized in energy. 
}
\label{figwf}
\end{figure}

In Fig. \ref{figsin1}, we present the numerical results for the transmission 
(top) through a Sinai billiard and for the phase rigidity  (bottom) 
as a function of energy $E$ and coupling strength $v$.
The shape of the billiard is shown in Figs. \ref{figwf}(a,b). 
Around $v\approx 1$, the transmission is enhanced 
and the phase rigidity  is reduced at all energies.
Analogous results (Fig. \ref{figsin2})  are obtained for the transmission
through a Sinai billiard with another geometry [Figs. \ref{figwf}(c,d)].
Comparing the two figures \ref{figsin1} and \ref{figsin2}, we see that
the  transmission spectrum as well as  the phase rigidity 
do change only a little when the geometry of the cavity is varied. 
This result coincides qualitatively
with the high reproducibility of the experimental data
observed in the fluctuations of the
spectrum of the atom $^{85}Rb$ in strong crossed magnetic and
electric fields \cite{stania}.

In Fig. \ref{figbun}, we show the numerical results for the transmission 
through a billiard of Bunimovich type and the phase rigidity  as a function
of energy and coupling strength. In the calculations, 
the leads are attached in such a manner, see Figs. \ref{figwf}(e,f), 
that the transmission via short-lived whispering gallery modes 
is supported by the geometry of the billiard in the regime of 
overlapping resonances.
The whispering gallery  modes are especially stable as has been
shown in earlier studies \cite{wgm} and as it can be seen directly by 
comparing Figs. \ref{figsin1}  and \ref{figsin2} with \ref{figbun}.

The correlation between the conductance $\langle G \rangle\, =\,
\langle |t|^2\rangle $ and 
$1-\langle |\rho|^2 \rangle $ can be seen more directly
from Fig. \ref{figcorr}, where $\langle G\rangle $ 
and $\langle |\rho|^2 \rangle $, averaged over energy, are shown 
as a function of $v$ for the three 
cavities whose geometries are shown in Figs. \ref{figwf}(a--f). 
In all cases, $\langle G\rangle $ and $1-\langle |\rho|^2 \rangle$ are 
strongly  correlated. Around $v=1$ 
$\langle G\rangle $ and  $1-\langle |\rho|^2  \rangle $ are maximum. 
Here also $N_E$, defined in (\ref{pricom2}), is maximum. 
When the parameters of the Sinai billiard
are changed from $x= 3, y=7$, Fig. \ref{figcorr}(a), to  $x=4, y=5$, 
Fig. \ref{figcorr}(b), $\langle G\rangle $ and 
$\langle |\rho|^2 \rangle $ do change only a little.

The situation is however another one for transmission through a Bunimovich 
cavity in the regime of overlapping resonances. 
When the leads are attached in such a manner that the transmission via 
whispering gallery modes is supported then 
the transmission is more enhanced and the phase rigidity is more reduced 
[Fig. \ref{figcorr}(c)] than in the  case of the two Sinai billiards 
considered above.
When the leads are attached in another manner to the Bunimovich cavity 
so that they do not support
transmission through whispering gallery modes, the transmission (averaged 
over energy) may be very small and, correspondingly, the phase rigidity 
(averaged over energy) remains large.  
For the geometry shown in Figs. \ref{figwf}(g,h), it is
$\langle G\rangle \; \le\; 0.1$ and 
$\langle |\rho|^2 \rangle \; \ge \; 0.9$. In this case, the 
whispering gallery modes are related, above all, to the reflection.

In Fig. \ref{figwf}, we show the wave functions $\phi_\lambda$
of a few short-lived and long-lived states in the four cavities considered. 
The corresponding eigenvalues $z_\lambda$ calculated at the energy $E$ 
are given in Table \ref{tab}. Although the  wave functions
are delocalized in space in all cases in the corresponding closed system,
the short-lived states might be localized  as can be seen from 
Figs. \ref{figwf}(a), (c), (e) and (g). Only the long-lived resonance
states are delocalized inside the cavity, 
see Figs. \ref{figwf}(b), (d), (f) and (h).

\begin{table}[h]
\caption{The eigenvalues $z_\lambda$ corresponding to the eigenfunctions 
$\phi_\lambda$ shown in Fig. \ref{figwf}, calculated at the energy $E$ and
coupling strength $v=1$.\\}

\begin{tabular}{ccc}
\hline\hline
~{Figure}~~ & \hspace*{1cm} E\hspace*{1cm} & \hspace*{1.5cm} $z_\lambda$ 
\hspace*{1.cm}\\
\hline
\ref{figwf}.a & 25 & $20.1228 - 3.1657~i$ \\
\ref{figwf}.c & 25 & $18.4570 - 2.3354~i$ \\
\ref{figwf}.e & 25 & $24.8227 - 2.4076~i$ \\
\ref{figwf}.g & 20 & $22.7141 - 1.1801~i$ \\
\hline
\ref{figwf}.b & 25 & $20.6891 - 0.0617~i$ \\
\ref{figwf}.d & 25 & $22.6227 - 0.0635~i$ \\
\ref{figwf}.f & 25 & $32.2046 - 0.0276~i$ \\
\ref{figwf}.h & 20 & $18.3281 - 0.0713~i$ \\
\hline\hline
\end{tabular}
\label{tab}
\end{table}

\section{Phase rigidity  and resonance trapping  
in the transmission through few-sites structures}

The results shown in
Fig. \ref{figcorr} show  some collective features  involved in the
scattering wave function $\Psi^E_C$. In order to understand these features
better, we provide  numerical results for simple structures 
with a  small number of sites. First, we consider
the one-dimensional case (chain) described by the tight-binding Hamilton
operator  
%-----------------------------------------------------------------(13)
\begin{eqnarray}
H=-\sum t_j|j\rangle\langle j+1|+c.c.
\label{chain1}
\end{eqnarray}
where $j$ runs over all sites of the system, i.e. over those 
of the left and right leads as well as over those of
the one-dimensional box consisting of $N$ sites. The $t_j$ are the hopping
matrix elements. Details of the model are given  in \cite{saro}.  
The box is opened by varying the hopping matrix elements
between leads and box in the following manner. We define $t_j=v_L$ if
$j=j_{\rm in}-1$; ~$t_j=v_R$ if $j=j_{\rm out}$; $t_j =1$
in the leads and $t_j=t_0$ otherwise. In the case $t_0=1$,
the eigenvalue spectrum $E_n=-\, 2\, t_0\, cos(\frac{n\pi}{N+1})$ of $H_B$
is distributed over almost the whole  propagation band $E(k) =-\, 2\, cos(k)$.
Our calculations are performed for  symmetrical cases with  
$v_L=v_R\equiv v$. The phase rigidity  is calculated from (\ref{rig2}).

\begin{figure}[t]
\includegraphics[width=.71\textwidth]{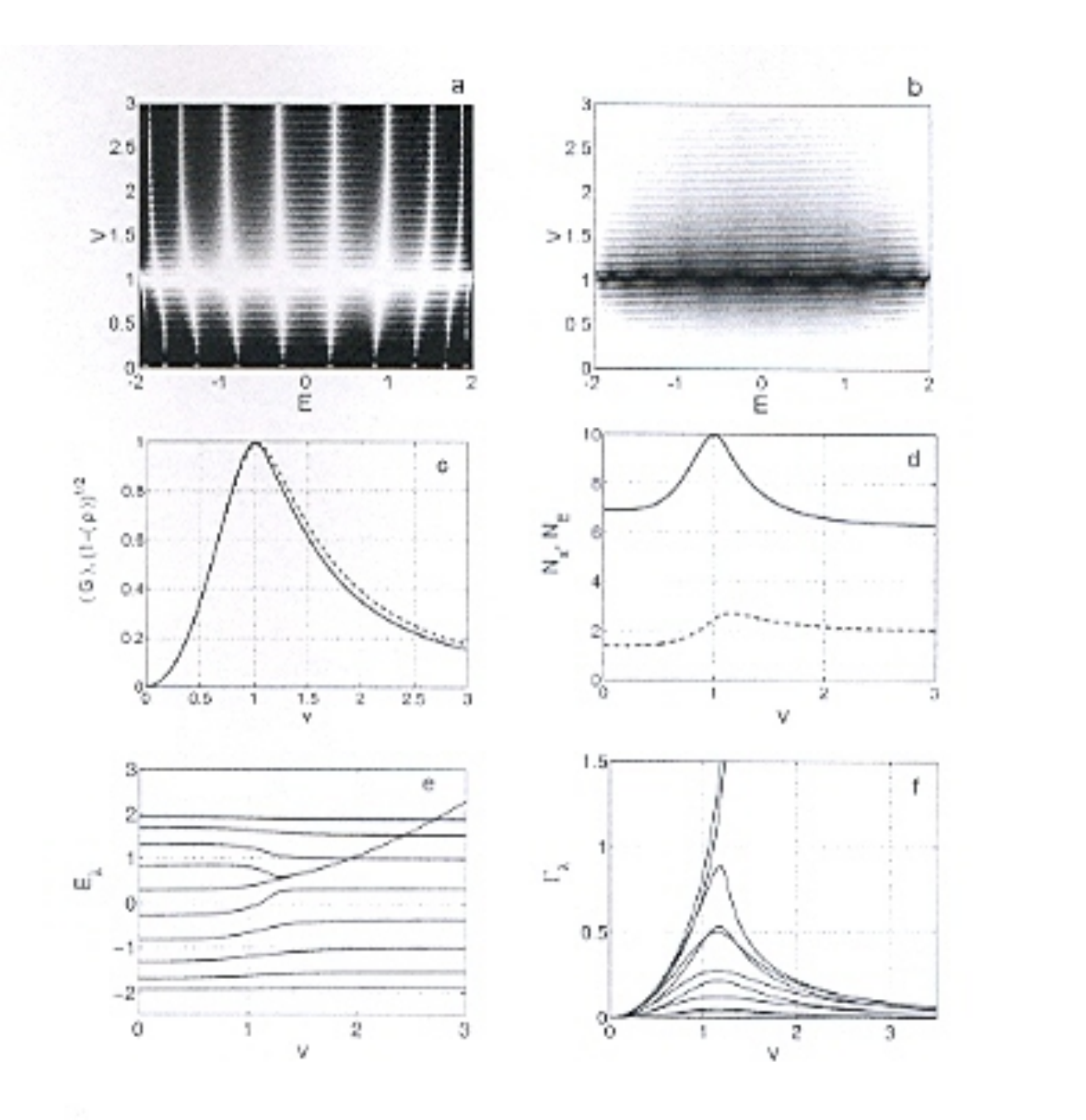}
\caption{
{\small
(a): The conductance $G=|t|^2$ of a one-dimensional 10-sites chain, 
$t_0=1$, with wires attached  at the ends of the chain
($j_{\rm in}=1, ~j_{\rm out}=10$) as a function of energy $E$ and
coupling strength $v$  between the chain and the two wires \cite{comm}.
The positions of the long-lived states at large coupling strength $v$ 
differ from the positions $E_\lambda^B$ (denoted by white circles). 
(b): The phase rigidity $|\rho|^2$ for the same 10-sites  chain. 
(c): Transmission $\langle G\rangle $ (full line) and 
$\sqrt{1-\langle |\rho|^2 \rangle}$ 
(dotted line), and  (d): $N_x$ (full line) and $N_E$ (dotted line), 
as a function of the coupling strength $v$
for the 10-sites chain. The values $\langle G\rangle $, 
$\langle |\rho|^2 \rangle $, $N_x$ and $N_E$ 
are averaged over energy in the whole energy window $-2 \le E \le 2$. 
(e) $E_\lambda$  and (f) $\Gamma_\lambda$ 
as  a function of the coupling strength $v$. The $E_\lambda$ and
$\Gamma_\lambda$ are calculated at $E=0.5$. 
All calculations are performed in the tight-binding model \cite{saro}.
}}
\label{figchain1}
\end{figure}

In Fig. \ref{figchain1}, 
we show results of calculations obtained  for a chain of length $M=10$
with $t_0=1$. In the sub-figure \ref{figchain1}(a,b) 
the conductance $G=|t|^2$ and the phase rigidity $|\rho|^2$ are  
shown as a function of  energy $E$ and coupling
strength $v$. These figures are very similar to Fig. \ref{figbun} for the
transmission through a quantum billiard with convex boundary 
in which the transmission occurs mainly via whispering gallery modes.
In the one-dimensional chain (Fig. \ref{figchain1}), we see threshold effects 
at $E=-2$ and $E=2$ where the widths of the resonance states
approach zero. These threshold effects
are excluded in Fig. \ref{figbun} where the energy range $13 \le E \le 34$
is considered and the thresholds are at about $E=10$ and 40. We conclude
therefore that the one-dimensional chain mimics the transmission through
the Bunimovich cavity occurring, at $v=1$, mainly via whispering gallery modes.
The transmission has a plateau at $v_{\rm cr} \approx 1$. In the case
of the Bunimovich cavity, dips appear due to
the long-lived trapped resonance states. 
Also the phase rigidity $|\rho|^2$ of the chain [Fig. \ref{figchain1}(b)]
shows a behavior similar to that obtained for the Bunimovich cavity. 
In the case of the chain, the transmission occurs clearly via traveling modes.

\begin{figure}[t]
\includegraphics[width=0.7\textwidth]{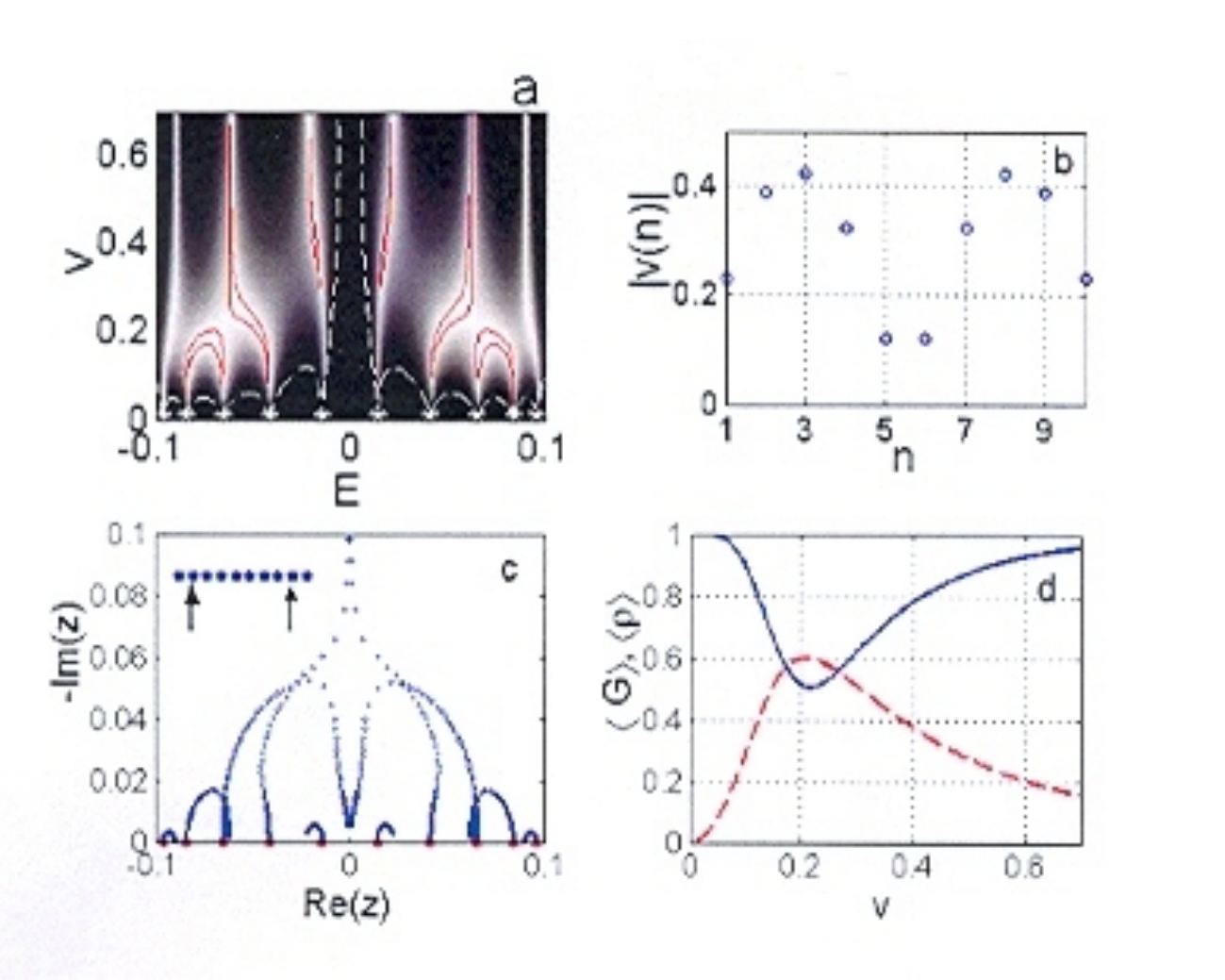}
\caption{{\small (color online)
The transmission through a one-dimensional chain of 10 sites, see  the 
inset of  sub-figure (c). The arrows show the sites at which the two 
(left and right) wires are attached. The hopping matrix elements inside the 
chain are 0.05 while those inside the attached wires are 1.
(a) Conductance $G$ versus incident energy $E$ and coupling matrix element
$v_L=v_R=v$, \cite{comm}. The white dashed lines show  regions of extremely 
low conductance (less than 0.02, corresponding to transmission gaps) 
while the red solid lines show  regions with enhanced 
transmission (more than 0.98, corresponding to  perfect conductance bands).
The positions of the bound states are shown by white stars.
(b) The coupling matrix elements $|V_n|=|\psi_n(2)|
=|\psi_n(9)|$ between the wires and 
the chain, $\psi_n(j)$, are the eigenfunctions of the "closed" chain described
by $H_B$, see Eq. (\ref{Heffgen}).
(c) Evolution of the complex eigenvalues $z_\lambda$
of the effective Hamilton operator,  Eq. (\ref{Heffgen}),
with increasing  coupling strength $v$. The $z_k$ are  calculated in
site representation according to Datta \cite{datta}. 
The values at $v=0$ are shown by red circles.
(d) Evolution of the mean conductance $\langle G\rangle$ (red broken line) 
and energy-averaged phase rigidity $\langle |\rho|^2 \rangle$ (blue full line) 
with $v$.  Note the hierarchical trapping of resonance states shown in (c):  
The resonance states 1 and 5 (10 and 6) are trapped already at small $v$. The
state 3 (8) first traps the state 2 (9) and becomes trapped by the state 4 (7)
at larger $v$. The two states 4 and 7 align with the two states in the two
channels. 
}}
\label{figchain2}
\end{figure}

\begin{figure}[t]
\includegraphics[width=0.7\textwidth]{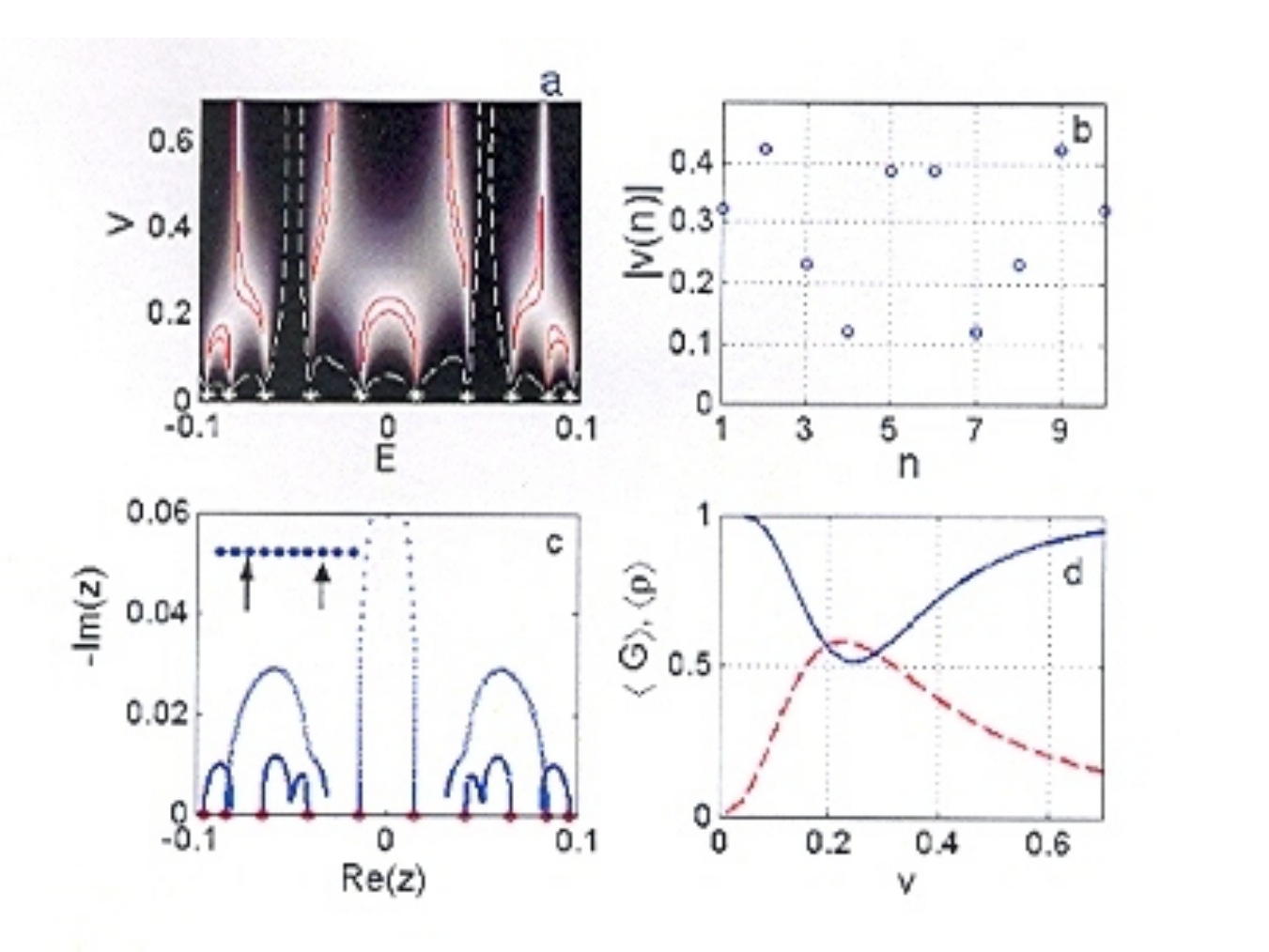}
\caption{
{\small (color online)
 The same as Fig. \ref{figchain2} but $|V_n|=|\psi_n(3)|
=|\psi_n(8)|$ in (b).
Note that resonance trapping shown in (c) starts at larger values of $v$ than
in Fig. \ref{figchain2}. Resonance state 2 (9) traps first the resonance
states 1, 3, 4 (10, 8, 7) and becomes trapped by state 5 (6) at larger $v$. 
The two states 5 and 6  in the center of the spectrum align with the two
states in the two channels.
}}
\label{figchain3}
\end{figure}

\begin{figure}[t]
\includegraphics[width=0.7\textwidth]{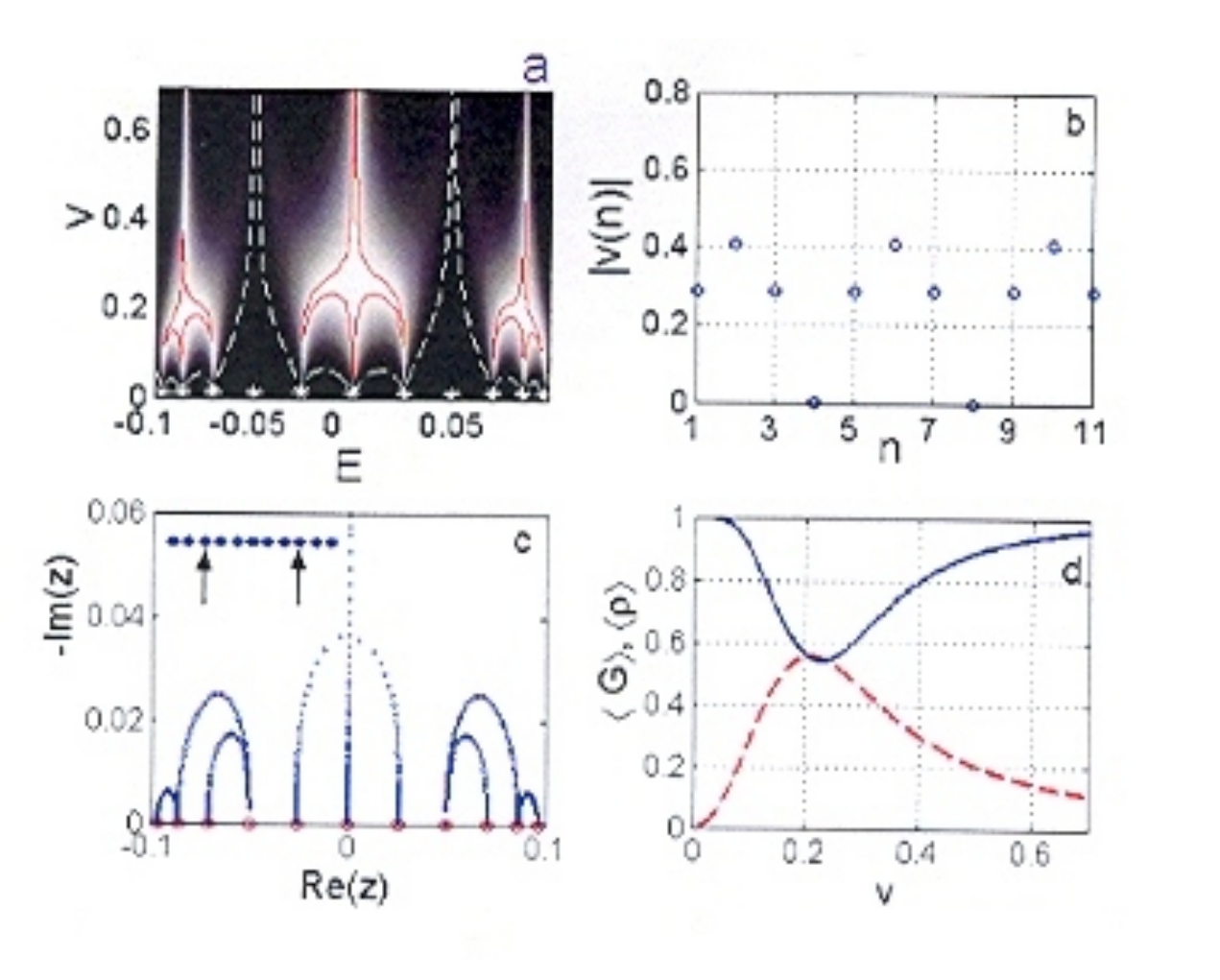}
\caption{
{\small (color online)
The same as Fig. \ref{figchain2} but for a chain of 11 sites and
$|V(n)|=|\psi_n(3)|= |\psi_n(9|$ in (b). 
Note that $|{\rm Im}(z_\lambda) |$ of the resonance state 4 (8) does not
increase with increasing $v$ even at small $v$, i.e. this state is trapped
already at very small $v$. The state 5 (or 7) traps first the states 
1, 2, 3 (11, 10, 9) and becomes trapped by the state 6 
(in the center of the spectrum)
at larger $v$. Finally, also in this case $|{\rm Im}(z_\lambda) |$ of 2 states
is large at large $v$, i.e. two states are aligned each to one of the states
in the two leads.
}}
\label{figchain4}
\end{figure}

\begin{figure}[t]
\includegraphics[width=0.7\textwidth]{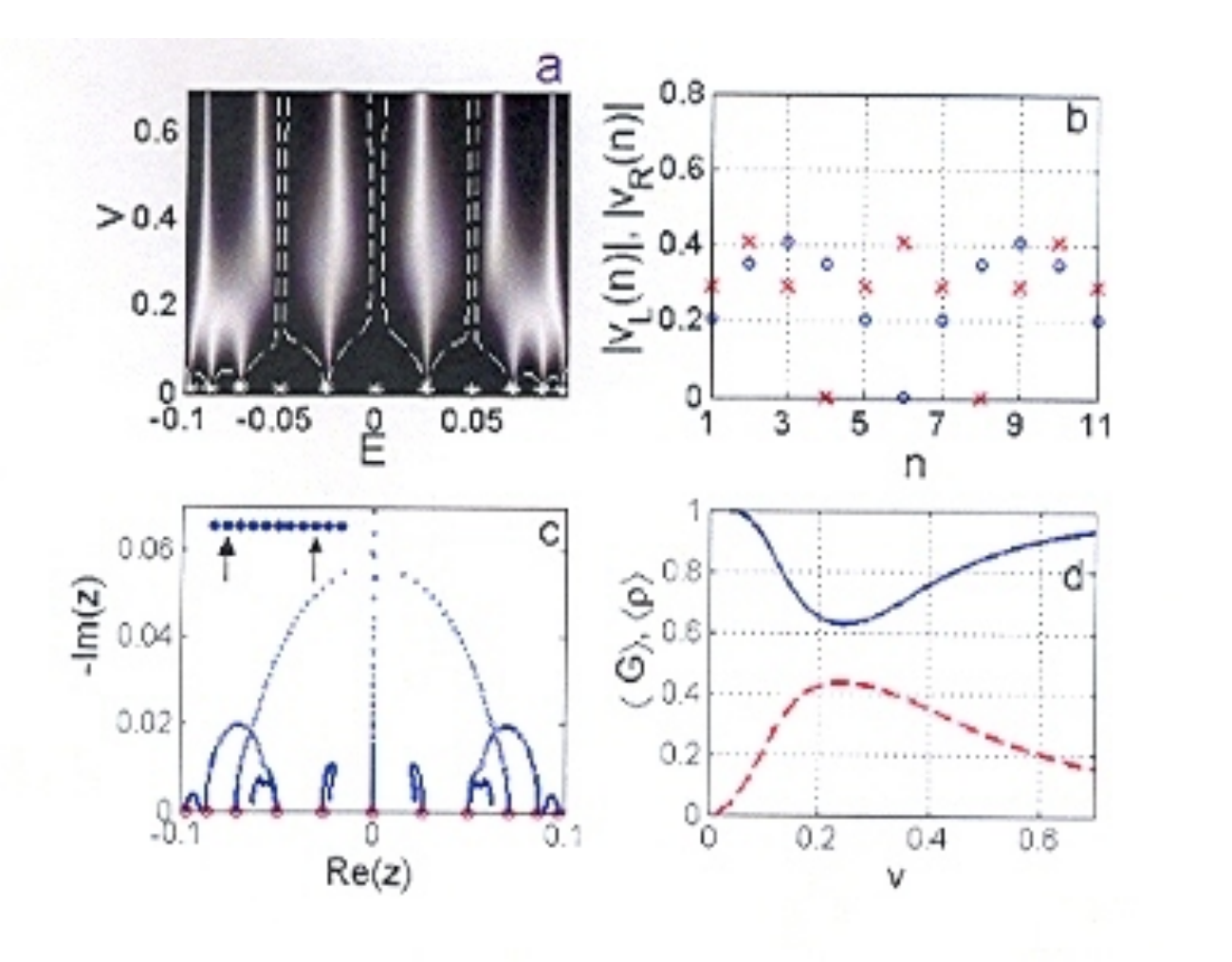}
\caption{
{\small (color online)
The same as Fig. \ref{figchain2} but for a chain of 11 sites and
$|V_L(n)|=|\psi_n(2)|$  (blue circles), $|V_R(n)|=|\psi_n(9)|$
(red crosses)  in (b). The  regions with enhanced 
transmission (more than 0.98, corresponding to  perfect conductance bands)
are small and not shown. 
Note the hierarchical resonance trapping also in this case. The
resonance states 1, 4, 5 (11, 8, 7) are trapped at relatively small $v$ while
the state 2 (10) first traps the neighboring states and 
then becomes trapped by the
state 3 (9) at larger $v$. With further increasing $v$, the state 3 (9)
becomes trapped by the state 6 in the center of the spectrum. Finally, there
are two states (6 and 3 or 9)
with large widths the wave functions of which are aligned each
to one of the two wave functions in the two leads. 
}}
\label{figchain5}
\end{figure}

For further discussion, we show additionally in Fig. \ref{figchain1} some
characteristic values discussed in the foregoing sections. 
Also in the one-dimensional chain,
we see the anti-correlation between conductance  $G=|t|^2$ and 
energy-averaged phase rigidity $|\rho|^2$ [Fig. \ref{figchain1}(c)]
that is observed in the billiards of Sinai and Bunimovich type
(Figs. \ref{figcorr}). The values $G=|t|^2$ and $\sqrt{1-|\rho|^2}$, 
averaged over energy, are  almost the same for all values $v$.
Furthermore,  $N_E$ is enhanced  in the critical region around $v\approx 1$
[Fig. \ref{figchain1}(d)]
in a similar manner as in Fig. \ref{figcorr} for the cavities.
Additionally, we show $N_x$ as a function $v$.
It is also enhanced in the critical region. 

In the chain, $N_x$ is smaller at $v\gg 1$ than at $v\ll 1$ 
according to the fact that the wave functions of the long-lived states 
appearing at large $v$ are, as a rule, more chaotic than the original ones
at small $v$ (see e.g. \cite{seligman}), i.e. they are distributed 
in $x$ with more or less equal probability over the whole cavity
in contrast to the original states at small $v$. 
As to $N_E$, it shows the opposite behavior, 
i.e. it is larger at $v\gg 1$ than at $v\ll 1$. The reason is that the
short-lived states are strongly correlated, i.e. the number $N_\lambda^p$
of principal components defined in (\ref{pricom1}), is large for every
short-lived state \cite{jung,ro91rep}. Due to this fact, $N_E$ remains 
comparably large when short-lived states are formed.

In the lower part of Fig. \ref{figchain1}, we show the  energies $E_\lambda$ 
and widths $\Gamma_\lambda$ of the 10 resonance states
as a function of the coupling strength $v$. We see the formation of
two short-lived states in the critical region $v \approx 1$
according to  the fact that two (identical) wires are attached.
The widths of all  states are maximal in the center of the energy window 
due to the condition $\Gamma_\lambda \to 0$ for all $\lambda$ in approaching
the thresholds $E \to \pm 2$. 
The short-lived states arise therefore from the middle of the
spectrum. Their contribution is shifted to  energies out of the window 
$-2 \le E \le 2$ when $E\ne 0$, see Fig. \ref{figchain1}(e)
calculated at $E=0.5$. We see therefore only 8 resonances in the transmission 
when $v>1$. Collective effects
are maximal around the critical value $v_{\rm cr}=1$ being somewhat smaller 
than the value  $v=1.2$ at which the widths bifurcate.
At  $v_{\rm cr}$,  many resonance states are almost aligned with the 
propagating modes $\xi^E_C$ in the leads, while at $v=1.2$
two resonance states are aligned at maximum.

In order to get a better understanding for the appearance of 
collective effects in the critical region around $v_{\rm cr}$ we show
in Figs. \ref{figchain2} to \ref{figchain5} the results of  
calculations performed for structures being  more complicated than a chain. 
We performed the calculations for 
different positions $j_{\rm in}$ and $j_{\rm out}$ at which the wires 
are attached, and for different numbers $M$ of resonance states. 
We choose $t_0=0.05$ (instead of $t_0=1$ in 
Fig. \ref{figchain1}). Due to the small value of $t_0$, the resonance
states lie in the narrow energy range $-0.1 \le E \le 0.1$, 
threshold effects are excluded and the eigenvalues $z_\lambda$
are almost independent of energy.

\begin{figure}[t]
\includegraphics[width=0.7\textwidth]{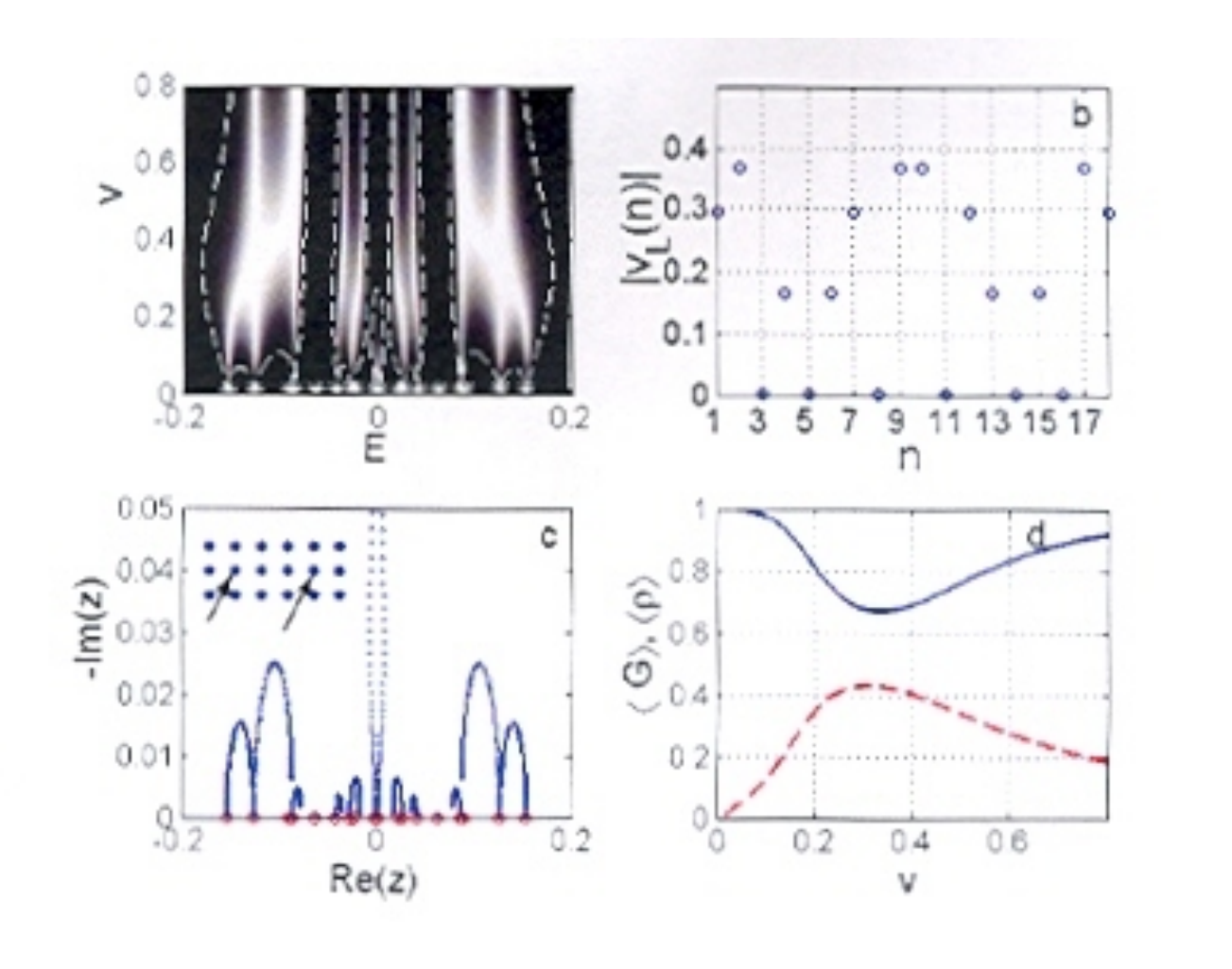}
\caption{
{\small (color online)
The same as Fig. \ref{figchain2} but for
a rectangle of 6 sites along 
the x-direction and 3 sites along the y-direction, see the inset in (c).
The coupling matrix elements $|V_L(m,n)|=|\psi_{m,n}(i=2,j=2)|, 
~|V_R(m,n)|=|\psi_{m,n}(i=5,j=2)|$ coincide because of the symmetry of the 
system, the $\psi_{m,n}(i,j)$ are the eigenfunctions of the "closed" 
rectangle  described by $H_B$, see Eq. (\ref{Heffgen}).
The  regions with enhanced 
transmission (more than 0.98, corresponding to  perfect conductance bands)
are small and not shown. 
Note that $|{\rm Im}(z_\lambda)|$ of the resonance state 4 (10) does not
increase with $v$ even at very small $v$. Resonance trapping occurs
hierarchically first around the center of the spectrum and then at its edges.
Finally, two resonance states align to the two scattering states in
the two leads.  
}}
\label{fig2d1}
\end{figure}

In the sub-figures (a) of  Figs. \ref{figchain2} to \ref{figchain5}, 
we show the transmission $G=|t|^2$ against coupling strength $v$ and energy $E$
for different one-dimensional systems. The areas with 
$0.98 \le |t|^2 \le 1$ are encircled by  red full lines and those with
$0 \le |t|^2 \le 0.02$ by white dashed lines. In these areas, the 
transmission at a fixed value $v$ shows a plateau of about 1 and 0,
respectively. 
While the positions in energy of the plateaus $|t|^2 \approx 1$ 
depend on $v$, the energies of the plateaus $|t|^2 \approx 0$ do almost 
not vary with $v$. They are corridor-like and divide the resonances
into different groups.  The number of corridors is equal to
$j_{\rm in}-1$ corresponding to an arrangement of the $M$ resonances in 
$j_{\rm in}$ groups.       
In each group,  plateaus $|t|^2 \approx 1$ appear in the critical region
of $v$, while 
the resonances cause transmission peaks of Breit-Wigner shape
at $v\ll v_{\rm cr}$ and   $v\gg v_{\rm cr}$.
The number of transmission peaks is equal to the number 
$M$ of resonance states when  $v\ll v_{\rm cr}$, but $M-2$ when 
$v\gg v_{\rm cr}$ due to the fact that two channels are open with  
which two of the resonance states align 
eventually in the critical region. 

In Figs. \ref{figchain2} to \ref{figchain4}, the wires  are attached 
symmetrically each to one of the internal sites, 
$j_{\rm out} = M - j_{\rm in}+1, ~j_{\rm in} >1$. 
In difference to the case  represented in Fig. \ref{figchain1},  
the incoming wave is split into two parts: both parts move 
in different directions each to one of the borders of the chain
where it  will be reflected. Then the two parts interfere and give rise to 
a complicated transmission picture. 
As a consequence of the interferences,  the 
transmission picture does almost not depend on the number of the sites
($M=10$ in Figs. \ref{figchain2} and \ref{figchain3}, but $M=11$ in Fig.
\ref{figchain4}). More important than the total number of sites 
is the distance  of the attached wires from the border of the chain.
When the distance is an odd number ($M_{\rm in} =2, ~M_{\rm out} = M-1$
in Fig. \ref{figchain2}) there is a zero transmission corridor
around $E=0$,  while the transmission is maximal  around this energy when 
the distance is an even number ($M_{\rm in} =3, ~M_{\rm out} = M-2$
in Figs. \ref{figchain3} and \ref{figchain4}).  The transmission
picture shows more structures when the symmetry is violated (Fig.
\ref{figchain5} for $M=11$ with $M_{\rm in} =2, ~M_{\rm out} = M-2$).

The regions   of extremely low conductance 
(less than 0.02, corresponding to transmission gaps) and  enhanced 
transmission (more than 0.98, corresponding to  perfect conductance bands),
marked in Figs. \ref{figchain2} to \ref{figchain4} 
by white dashed lines and red solid lines, respectively,  
are extended over comparably large energy regions. A similar 
result ("transmission gaps and sharp resonant states") has been found in
recent numerical calculations for the transport through a simple mesoscopic
device \cite{wahsh}. We underline that both the transmission plateaus 
with maximal transmission and the corridors of transmission zeros 
are the result of redistribution, interference and alignment processes. 
They can not be described by isolated Breit-Wigner resonances or
isolated  bound states in the continuum.
 
The redistribution processes taking place in the critical region around
$v_{\rm cr} = \sqrt{t_0}$   of the
coupling strength $v$ can be traced by comparing the sub-figures (b)
of Figs.  \ref{figchain2} to \ref{figchain5} with the corresponding 
sub-figures (c). In the sub-figures (b), the coupling coefficients $|v(n)|$
of the $M$ discrete states [eigenstates of $H_B$, Eq. (\ref{Heffgen})] 
to the propagating modes $\xi^E_C$ are shown.
They are different for the different positions $j_{\rm in}$
of attached wires and  different numbers $M$ of resonance states. 
Due to the symmetries involved in the considered chains, 
differences appear between the cases with an even and an odd 
number $M$ of resonance states. In the first case
[$M=10$, Figs. \ref{figchain2} and \ref{figchain3}], 
the coupling coefficients $v(n) ~(1\le n \le M)$ 
of all $M$ states to the continuum of propagating modes 
are different from
zero while there is at least one state with $v(n) =0$ in the last case
[$M=11$, Figs. \ref{figchain4} and \ref{figchain5}]. 
Nevertheless, the eigenvalue trajectories [sub-figures (c)] 
are similar for the different cases with different $M$.  
They show the typical resonance trapping behavior known  
from studies on many different systems, see  Ref. \cite{ro91rep}. 
In the regime of overlapping resonances
and two channels, the redistribution of the
spectroscopic properties of the system takes place due to the alignment of 
two resonance states each with one of the two 
propagating modes (input and output
channels). The redistribution occurs hierarchically \cite{isrodi,mudiisro}:
a resonance state will be trapped by overlapping with a neighboring state 
when its width is
somewhat smaller than that of the neighboring state. The widths of
trapped states cease from increasing with increasing coupling strength $v$.
Instead, they  mostly decrease  and the positions in energy 
of the trapped states do almost not change with further increasing $v$, 
for details see Ref. \cite{ro91rep}.  

The formation of 
zero-transmission corridors is directly related to the existence 
of trapped resonance states. They appear independently of the existence of
states whose coupling strength $v(n)$ is zero.  
Between the corridors, the redistribution processes create plateaus with an
enhanced transmission. These plateaus are related directly to the
avoided level crossings  appearing in the regime of overlapping resonances. 
Unlike the zero-transmission corridors,
the positions of the  plateaus with maximal transmission depend on the 
coupling strength $v$. 

In the sub-figures (d), the energy-averaged transmission 
$\langle G\rangle =\langle |t|^2\rangle $ and the
phase rigidity $\langle |\rho|^2\rangle $ 
of the total scattering wave function $\Psi^E_C$ are shown.
Like in the calculations for the microwave cavities,
the transmission is enhanced and the
phase rigidity  is reduced in the critical region 
where the spectroscopic redistribution takes place. 

The calculations for  two-dimensional systems are
performed in an analogous manner as those for one-dimensional systems, 
for details see Ref. \cite{saro}. The results 
are similar to those obtained for the one-dimensional systems. 
The main difference between the two cases  is that there are more 
possibilities to influence the transmission picture in the two-dimensional
systems than in the one-dimensional systems. An example is shown in Fig.
\ref{fig2d1}. 
It is interesting to see that the areas with enhanced transmission may be 
shifted and that the corridors with zero-transmission zeros may be 
relatively broad.

Summarizing the results shown in Figs. \ref{figchain1} to \ref{fig2d1}
for different toy models, we state the following. The enhancement of the
transmission  seen in realistic systems  in the crossover from the 
weak-coupling to the strong-coupling regime   (Sect. III)
is caused by spectroscopic redistribution processes taking place in the system
under the influence of the coupling to the continuum. 
These redistribution processes are correlated with a reduced value 
of the phase rigidity and  may be
traced back to the well-known resonance trapping phenomenon occuring in 
open quantum systems at high level density when the resonance states overlap.

\section{Fluctuation of the transmission probability}

\begin{figure}[t]
\includegraphics[width=.47\textwidth]{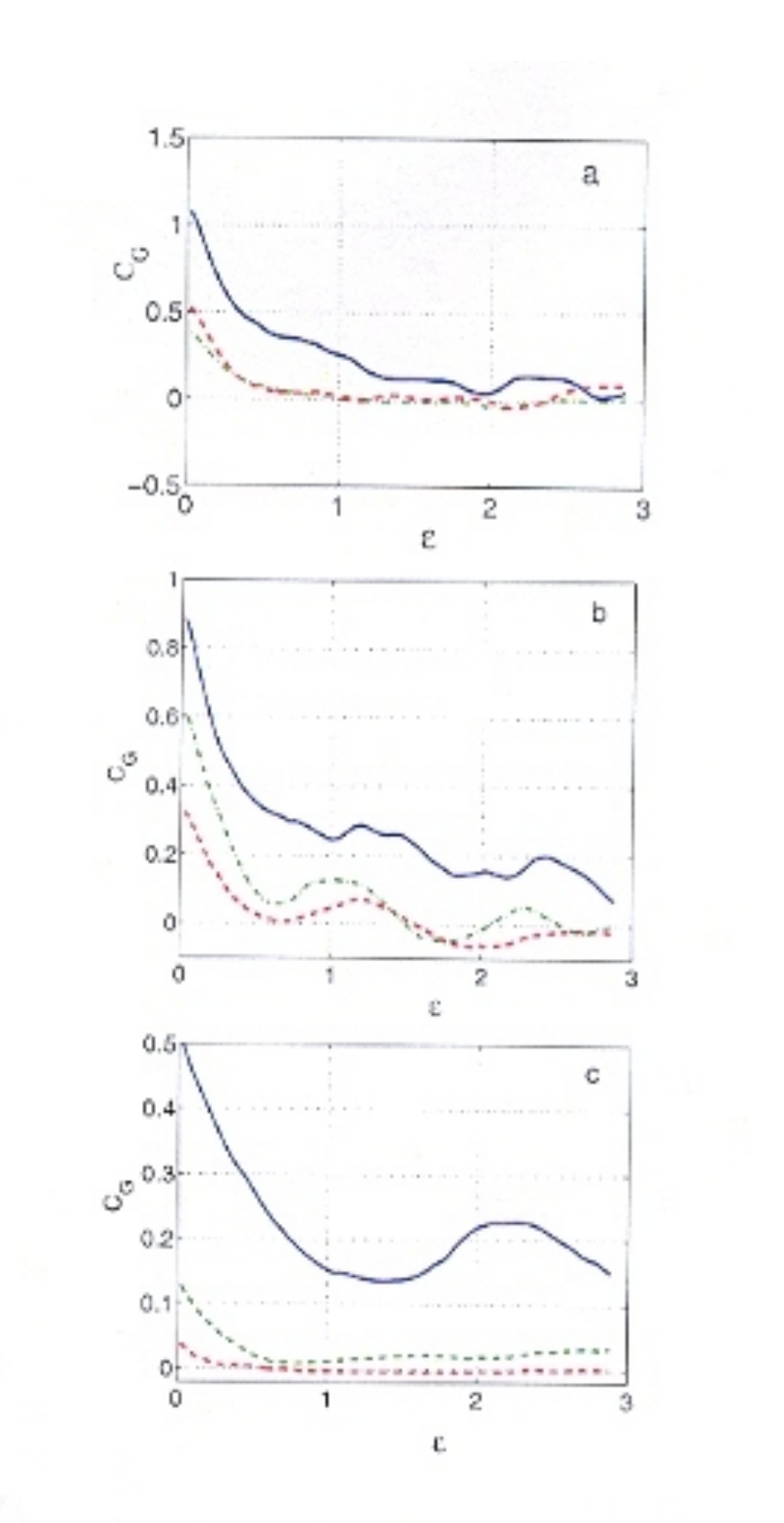}
\caption{\small (color online)
The autocorrelation function
$C_G(\varepsilon)$ for the transmission through 
the Sinai billiard with $x=3, y=7$ (a),  with $x=4, y=5$ (b)
and through the Bunimovich billiard  via whispering gallery modes (c).
In all three cases, the autocorrelation function is calculated for the 
whole energy window $11 \le E \le 34$ and shown for $v=0.9$ (red
dashed line), $v=1$ (blue full line) and $v=1.1$ (green dash-dotted line). 
}
\label{figfluc1}
\end{figure}

\begin{figure}[t]
\includegraphics[width=0.47\textwidth]{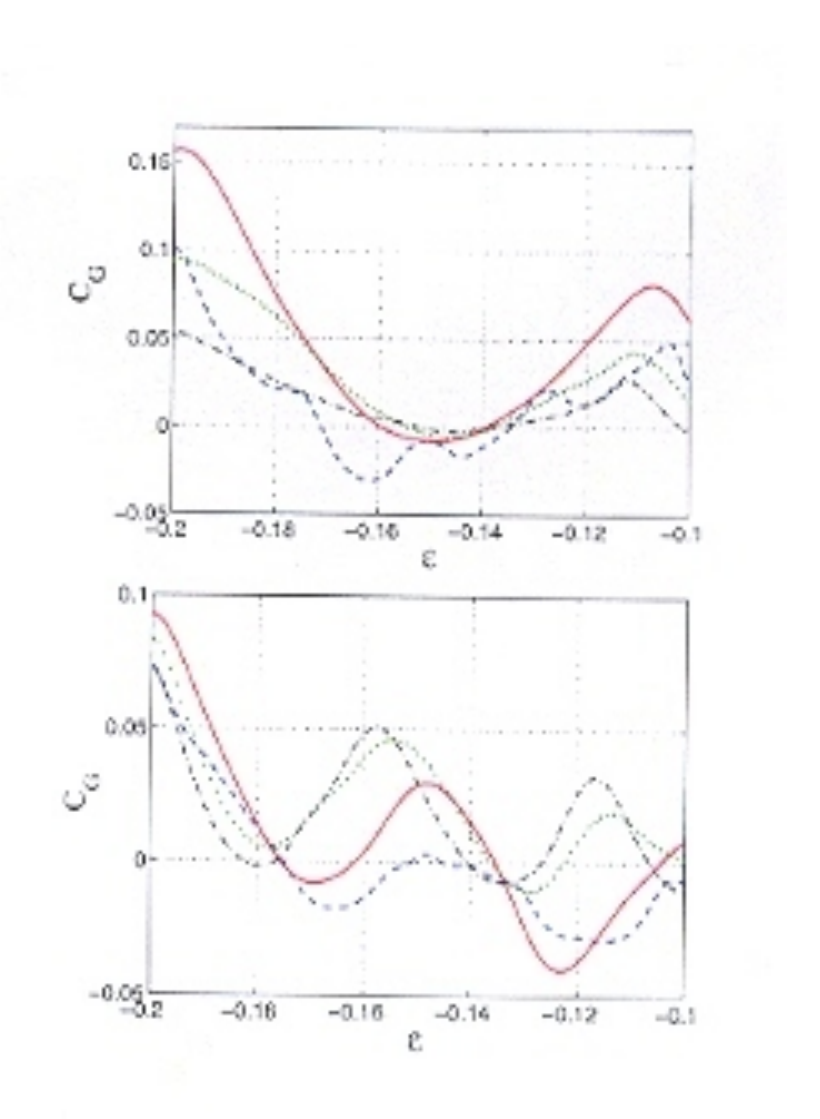}
\caption{ \small (color online)
The autocorrelation function
$C_G(\varepsilon)$ for the conductance $G=|t|^2$  of 
the  11-sites chain with, respectively, symmetrically (top) 
and non-symmetrically (bottom) attached leads [see 
the insets of the sub-figures (c) of Figs. \ref{figchain4} and 
\ref{figchain5}] for four different values of 
the coupling strength $v$: ~$v=v_{\rm cr}-0.1$ (dashed blue curves),
$v=v_{\rm cr}$ (solid red curves),
$v=v_{\rm cr}+0.1$ (dotted green curves), and
$v=v_{\rm cr}+0.2$ (dash-dotted black curves). 
}
\label{figfluc3}
\end{figure}

We performed an analysis of the transmission through the 
considered cavities  by means of the autocorrelation function 
of the transmission fluctuations 
$\delta  G = \delta |t|^2 = |t|^2 - \langle |t|^2\rangle$, 
see Eq. (\ref{eric1}). In the calculations, 
$I=23$ is the energy interval used to determine $\langle |t|^2\rangle $, 
i.e. $11 \le E \le 34$ as shown in Figs. \ref{figsin1}, \ref{figsin2} 
and \ref{figbun}. 
The area of all three cavities is of the same order of magnitude.
The number of states is about 35 in the considered energy region 
$11 \le E \le 34$. The average spacing  of the  states 
is  about 0.65. There is one channel in each of the two identical
attached leads. 

In Fig. \ref{figfluc1}, we show the autocorrelation function 
$C_G(\varepsilon)$ 
for the  transmission through the three cavities (considered in Sect. III)
in the regime of overlapping resonances with avoided level 
crossings where the transmission is enhanced and the phase rigidity 
is reduced. 
At small $\varepsilon$, we see the Lorentzian-like shape  characteristic   
of Ericson fluctuations \cite{ericson}. When $v=1$, the width of the  
Lorentzian is  larger than the average spacing. 
Comparing the function $C_G(\varepsilon)$ of the two Sinai billiards with that 
for the transmission through the
Bunimovich cavity via whispering gallery modes we see that 
the width of the Lorentzian is larger for the Bunivovich billiard 
than that for the two Sinai billiards.   
In all three cases, the width becomes smaller 
with decreasing as well as with increasing coupling strength, 
$v=0.9$ and 1.1, respectively. Enlarging $v$ further, we arrive in
the regime of separated time scales. Here,  $C_G(\varepsilon)$ falls off 
in a smaller region $\varepsilon$ than in the regime with overlapping
resonances. That means, the correlation between the long-lived states is small
at large $v$ and is comparably in value with the correlations between 
almost isolated resonance states (at small $v$).  

As can be seen from Fig. \ref{figfluc1},
the  function $C_G(\varepsilon)$ is still large at larger $\varepsilon$,  
especially in the case of the Bunimovich cavity. This result points to
long-range correlations existing between the different states,
above all between different  short-lived states. The oscillations differ  
from one another in the three considered cases. Related to this statement, 
we point to another result of our calculations:
in the autocorrelation function $C_G(\varepsilon)$ at $v\gg 1$, we have small
"peaks" (instead of oscillations) that are related to distant narrow resonance
states. 

In Fig. \ref{figfluc3}, 
the autocorrelation function $C_G$ versus
energy $E$ and coupling strength $v$ is shown for the transmission through 
the 11-sites chains, represented in the insets of the sub-figures (c) of 
Figs. \ref{figchain4} and \ref{figchain5}
with, respectively, symmetrically and non-symmetrically attached leads,
for four different values $v$ around the critical value $v_{\rm cr}$.
Both cases show the Lorentzian-like peak with maximum in the critical region 
$v=v_{\rm cr} =\sqrt{t_0} \approx 0.2$.  
The width of the peak increases with $v$ as long as  $v<v_{\rm cr} $.
This result agrees with theory \cite{ericson}.
Approaching however  $v_{\rm cr} $, the width of the Lorentzian is not well 
defined. For $v>v_{\rm cr} $, 
the mean width of the long-lived resonance states  
decreases as $v^{-2}$ (see e.g. Fig. \ref{figchain1}), while the width of the 
Lorentzian decreases much less.
The structure of the autocorrelation function is more pronounced in the 
non-symmetrical case than in the symmetrical one.
Altogether, the
results agree with those obtained for the Sinai and Bunimovich cavities
(Fig. \ref{figfluc1}).

The fluctuations in the transmission picture result, of course, 
from interferences.
However, the alignment of some of the wave functions with the propagating
modes in the wires plays an important role. 
It is expressed by the phase rigidity  as shown in Sects. III and IV.
The numerical results on the fluctuations demonstrate the weak dependence 
of the structure of the transmission picture
on external parameters such as the shape of the
billiard in Sect. III or the number of sites in the chains in Sect. IV.
They are a hint to the existence of collective effects, that are large
in the regime of overlapping resonances. 

We mention here the results of three completely different  experimental
studies that qualitatively agree with our theoretical results
on the collective effects  observed in the
conductance in the regime of overlapping resonances.  
First,  the directly measured average life time of compound nucleus
states in  $^{58,60}Ni(p,p')$ is significantly smaller than that of the
observed structures in the excitation function. Moreover, it  decreases with 
increasing bombarding energy of the projectile and is smaller than the average
distance between the states \cite{kanter}.
Secondly, a surprisingly high reproducibility of the fluctuation picture has
been observed in high-resolution experimental results on Ericson fluctuations
in atoms \cite{stania}. Thirdly,
long-range correlations between individual resonance states have
been found experimentally in an open microwave billiard \cite{brouwerkuhl}.

\section{Discussion of the results and summary}

In this paper, we studied the transmission through different 
mesoscopic systems with $M$ states  as a
function of the coupling strength $v$ between system and attached leads 
and wires, respectively. We considered the  two-channel case,
i.e. one propagating mode in each of the two  attached identical leads
(wires). The numerical calculations are performed by using the
Green function method (without any statistical assumptions).
The results show the following features. At small $v$, the transmission 
picture consists of $M$ isolated resonance peaks  
(corresponding to standing waves) while 
$M-2$ narrow resonances are superposed by a smooth "background" transmission 
(caused by the two traveling modes) at large $v$.  
Most interesting are the results in the regime
of strongly overlapping resonance states where the crossover from standing
to traveling waves takes place. Here, the transmission is enhanced.

According to our results, the
crossover occurs by means of the alignment of the wave functions of
the individual  resonance states each with one of the propagating modes
$\xi^E_{C=R}$ and  $\xi^E_{C=L}$
in the two leads.  The alignment  of the resonance states 
occurs stepwise. It is accompanied, at each step, by a certain
decoupling of at least one neighboring resonance state from the continuum of
propagating modes {\it (resonance trapping)}. 
For the mathematical foundation  of the alignment phenomenon in open quantum
systems see Ref. \cite{braro}.

Resonance trapping in the regime of overlapping resonance states  
occurs hierarchically as can be seen from Figs. \ref{figchain1} to 
\ref{fig2d1}. It is a collective effect to which, eventually, 
all the states in the considered energy region contribute. 
Therefore, also the reduction of the phase rigidity  
and  the enhancement of observable values,
such as the transmission, are  collective effects in the considered 
energy region. The reduction of the phase rigidity
and the enhancement of the transmission are 
maximal when a large number of individual resonance states is almost
aligned with the propagating modes in the leads. This happens at a
coupling strength $v$ being somewhat smaller than the critical one at which 
the number of aligned resonance states is exactly equal to the number of 
propagating modes (channels). This situation is illustrated best by
the whispering gallery modes appearing in a cavity of Bunimovich type,
Fig. \ref{figbun}, see also Figs.  \ref{figchain1} to \ref{fig2d1}.

Most interesting result of our numerical studies is that, 
in the regime of overlapping resonances, the conductance 
$\langle G\rangle $ is enhanced and correlated with 
$1-\langle |\rho|^2\rangle$.
For certain values of the coupling strength $v$ between system and environment,
the conductance  is plateau-like in some finite energy regions $\Delta E$. 
Furthermore, corridors with zero transmission
(total reflection) may appear as a function of the coupling
strength $v$ due to destructive interferences between neighboring resonance
states. The transmission picture with plateaus of maximal transmission
and  corridors 
of zero transmission is only weakly  influenced by the spectrum of the 
closed system. It results mainly from collective interference effects.

The relation of the  results obtained by us
to the idea of standing waves in almost closed systems and 
traveling waves in open systems \cite{shapiro}  is the following. 
In an almost closed system it is $H_{\rm eff} \approx H_B$, 
while in a strongly opened system the main term of $H_{\rm eff}$ is the 
coupling term $\sum_{C=L,R} V_{BC} (E^+-H_C)^{-1}V_{CB}$ via the continuum,
see Eq. (\ref{Heffgen}). The first case is realized in the regime of
non-overlapping resonances while in
the second case a smooth background (arising from the short-lived resonances) 
is superimposed by long-lived narrow (non-overlapping) resonances.
Furthermore, the monochromatic source considered in Ref. \cite{shapiro}
corresponds to the one-channel continuum represented by the
propagating modes  in  (\ref{Heffgen}). It is therefore not 
astonishing that the results obtained from  (\ref{trHeff}) for the 
transmission amplitudes in the two borderline cases 
fit well to the picture described 
in Ref. \cite{shapiro} for waves propagating in a random medium. 
However, the crossover between the two borderline cases is described 
differently in the two methods. In \cite{shapiro} a simple interpolation 
between the two  regimes is proposed.  In our formulation, however, the 
crossover between the two regimes is calculated in a straightforward
manner. The two equivalent exact expressions (\ref{trHeff}) and  (\ref{tr})
are used  for, respectively,  the two limiting cases and
the crossover regime with overlapping resonances. As a result of these
calculations, we received the result that the crossover 
regime is dominated by coherent collective phenomena.

The collective effects in the regime of overlapping resonance states       
influence also the autocorrelation function of the fluctuations of the
transmission probability. Long-range correlations do appear. The
autocorrelation function does not depend sensitively on external parameters
such as the shape of the Sinai billiard. 
The high reproducibility of the fluctuation picture observed 
experimentally on atoms \cite{stania} fits into this picture. 
Furthermore, our results coincide with those obtained in Ref. \cite{mois}:
in the few-channel case, Ericson-like fluctuations are determined by 
collective coherent  interferences between the resonance states.

Summarizing we state the following. The transition from the weak-coupling
to the strong-coupling regime is controlled by a redistribution of the
spectroscopic properties of the system under the influence of the 
environment to which the system is coupled (by attaching leads to it).
The redistribution consists in the alignment of 
a few resonance states each with one
of the propagating modes in the leads. The alignment is a  collective effect
to which all resonance states in the considered energy region contribute:
it occurs hierarchically by trapping  neighboring resonance states which
are more weakly coupled to the continuum of propagating modes
than those which align. Eventually, the number of aligned resonance states is
equal to the number of propagating modes in the leads. 
Due to their short life time (large decay width), the aligned resonance 
states form the background scattering on
which the trapped resonance states appear as narrow resonances. 
Since they are aligned  with the propagating modes $\xi^E_C$ in the leads, 
these short-lived resonance states may be identified with traveling modes.

{\acknowledgments}  
We thank the Max Planck Institute for the Physics of Complex Systems for
hospitality.


\begin{thebibliography}{99}
\bibitem{comm} The critical value $v_{\rm cr}$ of the coupling strength is
  determined by the ration $v/\kappa_0$ and $v/t_0$, respectively. In the
  figure, we show also results obtained numerically for $v>v_{\rm cr}$.
\end{thebibliography}
\end{document}